\documentclass{cernrep}
\usepackage{multirow}
\usepackage{textcomp}
\usepackage{hhline}
\usepackage{tabularx}
\usepackage[flushleft]{threeparttable}
\usepackage[table]{xcolor}
\usepackage{caption}
\usepackage{subcaption}
\usepackage{comment}
\usepackage{siunitx}

\newcommand{\ee}{$\rm{e}^+\rm{e}^-$}
\newcommand{\tabindent}{\hspace*{0.5cm}}

\usepackage{xcolor}
\definecolor{light-gray}{gray}{0.8}
\definecolor{apsblue}{rgb}{0.176, 0.152, 0.57}
\usepackage[colorlinks=true, linkcolor=black, citecolor=apsblue, filecolor=black, urlcolor=apsblue]{hyperref}

\begin{document}

\thispagestyle{empty}
\begin{centering}
{\Large {\bf HALHF: a hybrid, asymmetric, linear Higgs factory\\ using plasma- and RF-based acceleration}}\\
\end{centering}
\vspace{60pt}
\begin{figure}[h]
	\centering\includegraphics[width=0.2\textwidth]{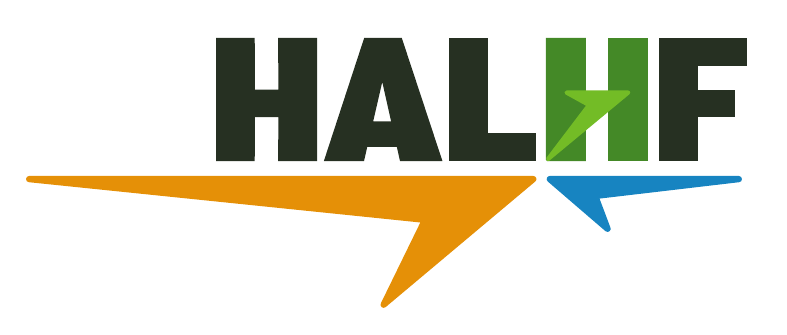}
\end{figure}
\vspace{180pt}
\begin{centering}
{\Huge {\bf Backup Document\\}}
\end{centering}
\vspace{260pt}
\hspace{-31pt}
\textit{Contact persons:} \\
Brian Foster, University of Oxford/DESY, brian.foster@physics.ox.ac.uk\\
Richard D'Arcy, University of Oxford, richard.darcy@physics.ox.ac.uk\\
Carl A. Lindstr{\o}m, University of Oslo, c.a.lindstrom@fys.uio.no\\

\newpage
\setcounter{page}{1}
\title{HALHF: a hybrid, asymmetric, linear Higgs factory\\ using plasma- and RF-based acceleration\\\textit{Backup Document}}
\author{The HALHF Collaboration: Erik Adli$^1$, Joshua Appleby$^2$, Timothy L. Barklow$^3$, Maria Enrica Biagini$^4$, Jonas Björklund Svensson$^5$, Mikael Berggren$^6$, Simona Bettoni$^7$, Stewart
Boogert$^8$, Philip Burrows$^2$, Allen Caldwell$^9$, Jian Bin Ben Chen$^1$, Vera Cilento$^{2,10}$, Laura Corner$^{11}$, Richard D’Arcy$^2$, Steffen Doebert$^{10}$, Wang Dou$^{12}$, Pierre Drobniak$^1$, Calvin Dyson$^{13}$, Sinead Farrington$^{14}$, John Farmer$^{9}$, Angeles Faus-Golfe$^{15}$, Manuel Formela$^6$, Arianne Formenti$^{16}$, Louis Forrester$^{13}$, Brian Foster$^{2,6}$, Jie Gao$^{12}$, Spencer Gessner$^3$, Niclas Hamann$^6$, Alexander 
Harrison$^2$, Mark J. Hogan$^3$, Eir Eline H{\o}rlyk$^1$, Maryam Huck$^6$, Daniel
Kalvik$^1$, Antoine
Laudrain$^6$, Reme Lehe$^{16}$, Wim Leemans$^6$, Carl A. Lindstr{\o}m$^1$, Benno List$^6$, Jenny List$^6$, Xueying Lu$^{17}$, Edward Mactavish$^{10}$, Vasyl Maslov$^6$, Emilio Nanni$^3$, John Osborne$^{10}$, Jens Osterhoff$^{16}$, Felipe Pe\~na$^{18,1}$,
Gudrid Moortgat Pick$^{19}$, Kristjan P\~{o}der$^6$, J\"{u}rgen Reuter$^6$,  Dmitrii Samoilenko$^6$, Nela Sedlackova$^{13}$, Andrei Seryi$^{20}$, Kyrre Sjobak$^1$, Terry Sloan$^{21}$, Steinar Stapnes$^{10}$, Rogelio Tomas Garcia$^{10}$, Maxim Titov$^{22}$, Malte Trautwein$^6$,  
Maxence Th\'evenet$^6$, Nicholas J. Walker$^6$, Marc Wenskat$^6$, Matthew Wing$^{23,6}$, Jonathan Wood$^6$.\\
\vspace{0.4cm}
$^1$ Department of Physics, University of Oslo, 0316 Oslo, Norway\\
$^2$ John Adams Institute for Accelerator Science at University of Oxford, Denys Wilkinson Building, Keble Road, Oxford, OX1 3RH, UK\\
$^3$ SLAC National Accelerator Laboratory, 2575 Sand Hill Road, CA 94025, Menlo Park, USA\\
$^4$ Laboratori Nazionali di Frascati, INFN, Via Enrico Fermi, 54, 00044 Frascati RM, Italy\\
$^5$ Department of Physics, Lund University, Box 118, 221 00 Lund, Sweden\\
$^6$ DESY, Notkestrasse 85, 22607, Hamburg, Germany\\
$^7$ Paul Scherrer Institute, Forschungsstrasse 111, 5232 Villigen, Switzerland\\ 
$^8$ Cockcroft Institute, Daresbury Laboratory, STFC, Keckwick Lane, Daresbury, WA4 4AD, Warrington, UK\\
$^{9}$ Max Planck Institut für Physik, Boltzmannstrasse 8, 85748 Garching/Munich, Germany\\
$^{10}$ CERN, CH-1211 Geneva 23, Switzerland\\
$^{11}$ Cockcroft Institute at University of Liverpool, School of Engineering, The Quadrangle, Brownlow Hill, Liverpool L69 3GH, UK\\
$^{12}$ Institute of High Energy Physics, Chinese Academy of Sciences, 9 Yuquan Rd, Shi Jing Shan Qu, Bei Jing Shi, 100039, China.\\ 
$^{13}$ John Adams Institute for Accelerator Science at Department of Physics, Imperial College London, Prince Consort Road, South Kensington, London SW7 2BW, UK\\
$^{14}$ Rutherford Appleton Laboratory, STFC, Harwell Campus, OX11 0QX, Didcot, UK\\
$^{15}$ Laboratoire de Physique des 2 Infinis Irène Joliot-Curie, IJCLab, Orsay, Bât. 100 et 200, 15 rue Georges Clémenceau, F-91405 Orsay, France\\
$^{16}$ Lawrence Berkeley National Laboratory, 1 Cyclotron Rd, Berkeley, CA 94720, USA\\
$^{17}$ Argonne National Laboratory, 9700 S Cass Avenue, IL60439, Lemont, USA\\
$^{18}$ Ludwig-Maximilians-Universität München, Geschwister-Scholl-Platz 1, 80539 München, Germany \\
$^{19}$ II. Institute of Theoretical Physics, University of Hamburg, Luruper Chaussee 149, 22761 Hamburg, Germany \\
$^{20}$ Thomas Jefferson National Accelerator Facility, 12000 Jefferson Avenue, VA23606, Newport News, USA\\
$^{21}$ University of Lancaster, Bailrigg, Lancaster LA1 4YW, UK\\
$^{22}$ CEA-IRFU, Bât. 141 CEA - Saclay, 91191 Gif-sur-Yvette, France\\
$^{23}$ Department of Physics \& Astronomy, University College London, Gower St, London WC1E 6BT, UK}
\maketitle
\tableofcontents
\clearpage
\section{Introduction}
\label{sec:intro}
This document expands on the Comprehensive Summary submitted to the EPPSU 2026~\cite{Input2025}. It contains details on aspects of the HALHF project \cite{HALHF,HALHF_upgrades} that could not be fitted into the Summary. Some sections contain work that is still preliminary and/or status reports on current progress.

\section{Overall design}

\subsection{Layout schematics}
\label{sec:schematics}
\begin{figure}[h]
    \centering
	\includegraphics[width=0.855\textwidth]{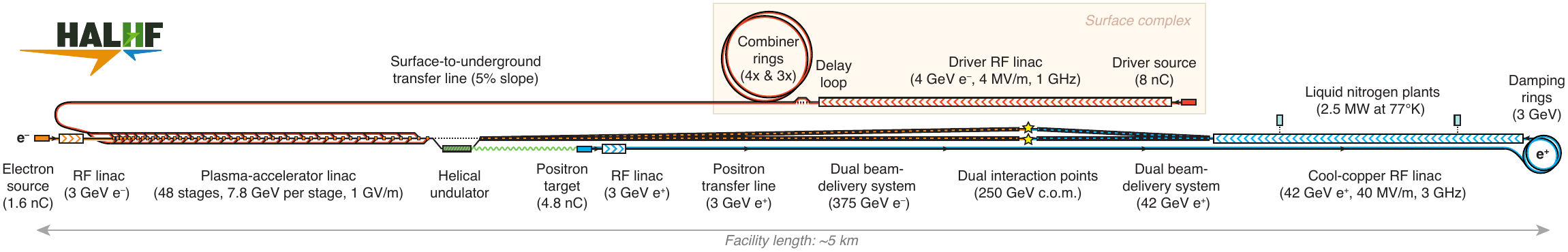}
    \caption{Schematic view of the new HALHF baseline at 250~GeV CoM. The red sections relate to driver electrons, orange to colliding electrons, blue to positrons and green to photons. Other components are as labelled on the figure.}
    \label{Fig-1}
\end{figure}

\begin{figure}[h]
    \centering
    \includegraphics[width=\textwidth]{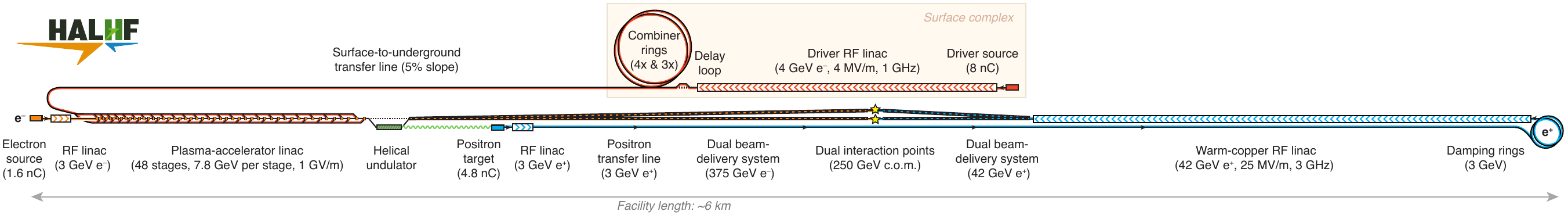}
    \caption{Schematic view of HALHF with the fall-back warm RF positron linac. Other details as in the caption to Fig.~\ref{Fig-1}.}
    \label{Fig-2}       
\end{figure}

Figure~\ref{Fig-1} and ~\ref{Fig-2} show two layouts of HALHF. The first is the preferred baseline, in which the positron linac is provided by cool copper technology with a gradient of 40~MV/m. Figure~\ref{Fig-2} shows an alternative layout using more conventional but lower gradient warm technology, based on the SLAC linac (here assuming 25~MV/m).  

\subsection{Parameter optimisation process}
\label{sec:parameteroptimisation}
The new baseline of HALHF (i.e., ``HALHF 2.0"), as described in Ref.~\cite{Foster2025}, was produced from a detailed process of optimisation using a Bayesian optimiser with inputs for costs of individual parameters derived from experience and a wide variety of background literature. The input cost basis is given in detail in Sect.~\ref{sec:costbasis}. 

\begin{table*}[p]
    \tiny
    \centering
    \begin{tabular}{p{0.28\linewidth}>{\centering}p{0.06\linewidth}>{\centering}p{0.08\linewidth}>{\centering\arraybackslash}p{0.08\linewidth}>{\centering\arraybackslash}p{0.08\linewidth}>{\centering\arraybackslash}p{0.08\linewidth}>{\centering\arraybackslash}>{\centering\arraybackslash}p{0.08\linewidth}>{\centering\arraybackslash}p{0.08\linewidth}}
    
    \textit{Machine parameters} & \textit{Unit} & 
   \multicolumn{2}{c}{\textit{Value (250 GeV)}}& \multicolumn{2}{c}{\textit{Value (380 GeV)}} & \multicolumn{2}{c}{\textit{Value (550 GeV)}}\\
    \hline
    Centre-of-mass energy & GeV & \multicolumn{2}{c}{250} & \multicolumn{2}{c}{380} &\multicolumn{2}{c}{550}\\
    Centre-of-mass boost & ~ & \multicolumn{2}{c}{1.67} & \multicolumn{2}{c}{1.67} & \multicolumn{2}{c}{1.67}\\
    Bunches per train & & \multicolumn{2}{c}{160} & \multicolumn{2}{c}{160}  & \multicolumn{2}{c}{160}  \\
    Train repetition rate & Hz & \multicolumn{2}{c}{100} & \multicolumn{2}{c}{100} & \multicolumn{2}{c}{100}\\
    Average collision rate & kHz & \multicolumn{2}{c}{16} & \multicolumn{2}{c}{16} & \multicolumn{2}{c}{16}\\
    Luminosity & cm$^{-2}$~s$^{-1}$ & \multicolumn{2}{c}{$1.2\times10^{34}$} & \multicolumn{2}{c}{$1.7\times10^{34}$} & \multicolumn{2}{c}{$2.5\times10^{34}$} \\
    Luminosity fraction in top 1\% & & \multicolumn{2}{c}{63\%} & \multicolumn{2}{c}{53\%} &\multicolumn{2}{c}{46\%} \\
    Quantum parameter ($\Upsilon$) & & \multicolumn{2}{c}{0.9} & \multicolumn{2}{c}{1.6} &\multicolumn{2}{c}{2.8} \\
    Estimated total power usage & MW & \multicolumn{2}{c}{106} & \multicolumn{2}{c}{154} & \multicolumn{2}{c}{218} \\
        Total site length & km & \multicolumn{2}{c}{4.9} & \multicolumn{2}{c}{6.5} & \multicolumn{2}{c}{8.4} \\
    \hline
     & & & \\[-6pt]
    \textit{Colliding-beam parameters} & & $e^{-}$ & $e^{+}$ & $e^{-}$ & $e^{+}$ & $e^{-}$ & $e^{+}$ \\ 
    \hline
    Beam energy & GeV & 375 & 41.7 & 570 & 63.3 & 825 & 91.7 \\
    Bunch population & $10^{10}$ & 1 & 3 & 1 & 3 & 1 & 3 \\
    Bunch length in linacs (rms) & $\mu$m & 40 & 150 & 40 & 150 & 40 & 150 \\ 
    Bunch length at IP (rms) & $\mu$m & \multicolumn{2}{c}{150} & \multicolumn{2}{c}{150} & \multicolumn{2}{c}{150}  \\
    Energy spread (rms) & \% & \multicolumn{2}{c}{0.15} & \multicolumn{2}{c}{0.15} & \multicolumn{2}{c}{0.15} \\
    Horizontal emittance (norm.) & $\mu$m & 90 & 10 & 90 & 10 & 90 & 10 \\
    Vertical emittance (norm.) & $\mu$m & 0.32 & 0.035 & 0.32 & 0.035 & 0.32 & 0.035\\
    IP horizontal beta function & mm & \multicolumn{2}{c}{3.3} & \multicolumn{2}{c}{3.3} & \multicolumn{2}{c}{3.3}\\
    IP vertical beta function & mm & \multicolumn{2}{c}{0.1} & \multicolumn{2}{c}{0.1} & \multicolumn{2}{c}{0.1} \\
    IP horizontal beam size (rms) & nm & \multicolumn{2}{c}{636} & \multicolumn{2}{c}{519} & \multicolumn{2}{c}{429} \\
    IP vertical beam size (rms) & nm & \multicolumn{2}{c}{6.6} & \multicolumn{2}{c}{5.2} & \multicolumn{2}{c}{4.4} \\
    Average beam power delivered & MW & 9.6 & 3.2 & 14.6 & 4.9 & 21.1 & 7.0 \\
    Bunch separation & ns & \multicolumn{2}{c}{16} &\multicolumn{2}{c}{16} &\multicolumn{2}{c}{16} \\
    Average beam current & {\textmu}A & 26 & 77 & 26 & 77 & 26 & 77  \\[2pt]
    \hline \\[-6pt]
    \multicolumn{2}{l}{\textit{Positron cool-copper RF linac parameters (S-band)}}  & & & & \\
     \hline
     Average cavity gradient & MV/m & \multicolumn{2}{c}{40}& \multicolumn{2}{c}{40} & \multicolumn{2}{c}{40}\\
        Average gradient & MV/m & \multicolumn{2}{c}{36} & \multicolumn{2}{c}{36} & \multicolumn{2}{c}{36}\\
    Wall-plug-to-beam efficiency & \% & \multicolumn{2}{c}{11} & \multicolumn{2}{c}{11} & \multicolumn{2}{c}{11}\\
    RF power & MW & \multicolumn{2}{c}{11.7} &\multicolumn{2}{c}{17.8} &\multicolumn{2}{c}{25.8} \\
    Cooling power  & MW & \multicolumn{2}{c}{17.9} &  \multicolumn{2}{c}{27.3} & \multicolumn{2}{c}{39.5}\\
     Total power & MW & \multicolumn{2}{c}{29.6} & \multicolumn{2}{c}{45.1} & \multicolumn{2}{c}{65.3} \\
   Klystron peak power & MW & \multicolumn{2}{c}{67} & \multicolumn{2}{c}{67} & \multicolumn{2}{c}{67} \\
     Number of klystrons & ~ & \multicolumn{2}{c}{321} & \multicolumn{2}{c}{452} & \multicolumn{2}{c}{678} \\
    RF frequency & GHz & \multicolumn{2}{c}{3} & \multicolumn{2}{c}{3} & \multicolumn{2}{c}{3} \\
 Operating Temperature & K & \multicolumn{2}{c}{77} & \multicolumn{2}{c}{77} & \multicolumn{2}{c}{77} \\
         Length (after damping ring, starting at 3 GeV) & km & \multicolumn{2}{c}{1.1} & \multicolumn{2}{c}{1.7} & \multicolumn{2}{c}{2.5} \\
    \hline \\[-6pt]
    \multicolumn{2}{l}{\textit{Driver linac RF parameters (L-band)}} & & & & \\
    \hline \\[-6pt]
        Average cavity gradient & MV/m & \multicolumn{2}{c}{4} & \multicolumn{2}{c}{4} & \multicolumn{2}{c}{4}\\
    Average gradient & MV/m & \multicolumn{2}{c}{3} & \multicolumn{2}{c}{3} & \multicolumn{2}{c}{3} \\
    Wall-plug-to-beam efficiency & \% & \multicolumn{2}{c}{55} & \multicolumn{2}{c}{55} & \multicolumn{2}{c}{55} \\
    RF power usage & MW & \multicolumn{2}{c}{42.9} & \multicolumn{2}{c}{66.0} & \multicolumn{2}{c}{96.4}\\
   Klystron peak power & MW & \multicolumn{2}{c}{21} &  \multicolumn{2}{c}{21} & \multicolumn{2}{c}{21}\\
     Number of klystrons & ~ & \multicolumn{2}{c}{409} & \multicolumn{2}{c}{630} & \multicolumn{2}{c}{919} \\
    RF frequency & GHz & \multicolumn{2}{c}{1} & \multicolumn{2}{c}{1} & \multicolumn{2}{c}{1} \\
        Length & km & \multicolumn{2}{c}{1.3} & \multicolumn{2}{c}{1.9} & \multicolumn{2}{c}{2.8} \\
\hline \\[-6pt]
\multicolumn{2}{l}{\textit{Combiner Ring parameters}} 
    & & \\
    \hline \\[-6pt]
    ~ & ~ & ~ & ~ & ~ & ~ \\[-8pt]
    Delay loop length & m & \multicolumn{2}{c}{1.5} &\multicolumn{2}{c}{1.5} &\multicolumn{2}{c}{1.5} \\
    CR1 diameter & m & \multicolumn{2}{c}{244} &\multicolumn{2}{c}{244} &\multicolumn{2}{c}{244} \\
    CR2 diameter & m & \multicolumn{2}{c}{244} &\multicolumn{2}{c}{244} &\multicolumn{2}{c}{244} \\
    \hline \\[-6pt]
\multicolumn{2}{l}{\textit{PWFA linac and drive-beam parameters}} 
    & & & & \\
    \hline \\[-6pt]
    Number of stages &  & \multicolumn{2}{c}{48} & \multicolumn{2}{c}{48} & \multicolumn{2}{c}{48} \\
    Plasma density & cm$^{-3}$ & \multicolumn{2}{c}{$6\times10^{14}$} & \multicolumn{2}{c}{$6\times10^{14}$} & \multicolumn{2}{c}{$6\times10^{14}$}\\ 
    In-plasma accel. gradient & GV/m & \multicolumn{2}{c}{1}& \multicolumn{2}{c}{1} & \multicolumn{2}{c}{1} \\
    Av. gradient (incl. optics) & GV/m & \multicolumn{2}{c}{0.33}& \multicolumn{2}{c}{0.38} & \multicolumn{2}{c}{0.43} \\
    Transformer ratio & ~ & \multicolumn{2}{c}{2} & \multicolumn{2}{c}{2} & \multicolumn{2}{c}{2} \\
    Length per stage & m & \multicolumn{2}{c}{7.8} &\multicolumn{2}{c}{11.8} & \multicolumn{2}{c}{17.1} \\
    Energy gain per stage\tnote{a} & GeV & \multicolumn{2}{c}{7.8} & \multicolumn{2}{c}{11.8} & \multicolumn{2}{c}{17.1} \\
    Initial injection energy & GeV & \multicolumn{2}{c}{3} & \multicolumn{2}{c}{3} & \multicolumn{2}{c}{3}\\
    Driver energy & GeV & \multicolumn{2}{c}{4}  & \multicolumn{2}{c}{5.9} & \multicolumn{2}{c}{8.6}\\
    Driver bunch population & $10^{10}$ & \multicolumn{2}{c}{5.0} & \multicolumn{2}{c}{5.0} & \multicolumn{2}{c}{5.0} \\
    Driver bunch length (rms)  & $\mu$m &  \multicolumn{2}{c}{253} &  \multicolumn{2}{c}{253} & \multicolumn{2}{c}{253} \\
    Driver average beam power & MW & \multicolumn{2}{c}{23.8} & \multicolumn{2}{c}{36.2} & \multicolumn{2}{c}{52.6} \\
    Driver bunch separation & ns & \multicolumn{2}{c}{4} & \multicolumn{2}{c}{4} & \multicolumn{2}{c}{4} \\
    Driver-to-wake efficiency & \% & \multicolumn{2}{c}{80} & \multicolumn{2}{c}{80} & \multicolumn{2}{c}{80} \\
    Wake-to-beam efficiency & \% & \multicolumn{2}{c}{50} & \multicolumn{2}{c}{50} & \multicolumn{2}{c}{50} \\
    Driver-to-beam efficiency & \% & \multicolumn{2}{c}{40} & \multicolumn{2}{c}{40} & \multicolumn{2}{c}{40} \\
    Wallplug-to-beam efficiency & \% & \multicolumn{2}{c}{22} & \multicolumn{2}{c}{22} & \multicolumn{2}{c}{22} \\
    Cooling req.~per stage length & kW/m & \multicolumn{2}{c}{38.4} & \multicolumn{2}{c}{38.4} & \multicolumn{2}{c}{38.4} \\
    Length & km & \multicolumn{2}{c}{1.1} & \multicolumn{2}{c}{1.5} & \multicolumn{2}{c}{1.9} \\
    \hline
    \end{tabular}
   \caption{HALHF parameters for the updated baseline design at 250~GeV, 380~GeV and and 550~GeV CoM energies. (This table is reproduced from Table 1 in the Comprehensive Summary~\cite{Input2025}.)}
   \label{tab:1}
\end{table*}

In order to carry out the optimisation, a new ``system code" (ABEL) was developed: this simultaneously performed start-to-end physics simulations and produced engineering layouts and cost estimates based on the physics performance. Specifically, the Bayesian optimisation process (using the Ax framework \cite{Ax}) used 12 parameters: the energy asymmetry; the number of bunches in a train; the repetition rate of the trains; the combiner-ring compression factor; the drive-bunch temporal separation; the number of RF cells per structure in the driver linac; the number of structures per klystron in the driver linac; the accelerating gradient in the driver linac; the accelerating gradient in the positron linac; the number of RF cells per klystron in the positron linac; the number of PWFA stages; and the PWFA transformer ratio. The remaining parameters were calculated based on these numbers. The PWFA gradient was set to 1~GV/m and plasma density to 6$\times$10$^{14}$~cm$^{-3}$ prior to optimisation in order to ensure a conservative design. Approximately 80 iterations of the Bayesian optimisation were sufficient to locate the global optimum (this optimisation was repeated several times with similar results). The resulting parameter set for HALHF produced by this optimisation process is shown in Table~\ref{tab:1}.

\section{Cost estimate basis}
\label{sec:costbasis}

This section discusses the detailed costing of the construction (for direct comparison to other collider proposals), as well as additional costs used in the Bayesian optimisation process. For convenience of comparison, all costs not originally expressed in ILC cost units (ILCU; i.e., 2012 dollars) have been scaled to ILCUs and subsequently converted to other currency units as required using the appropriate official deflators.

\subsection{Construction costs}
The costing was performed using the newly developed system code ABEL (which will be made publicly available soon). It performs start-to-end simulations of the collider, where each element used is costed based on its most salient characteristic, \textit{viz.} length, volume, power or individually per component: the per-element costs are summarised in Table~\ref{tab:cost}. The resulting overall costs are listed in Table 4 in the Comprehensive Summary~\cite{Input2025}.

\begin{table}[p]
\scriptsize
    \begin{tabular}{p{0.23\linewidth}>
    {\centering}p{0.09\linewidth}>
    {\centering}p{0.07\linewidth}>
    {\centering}p{0.07\linewidth}>
    {\centering}p{0.07\linewidth}>
    {}p{0.03\linewidth}>
    {}p{0.25\linewidth}}%
    \hline
    
    {\bf Cost element (per length)}&Cost/length & \multicolumn{3}{c}{Length (m)}  & Ref. & Comment \cr
    & (kILCU/m) & 250 GeV  & 380 GeV  & 550 GeV & &  \cr
    \hline
    Accelerating structures & 115.00 & 2,052 & 3,102 & 4,474 & CLIC & Assumed same for L- \& S-band. \cr
    Damping rings & 260.00  & 767 & 767 & 767 & CLIC & Two rings in one tunnel. \cr
    Combiner ring & 79.00  & 1,535 & 1,535 & 1,535 & CLIC & Two rings in one tunnel. \cr
    Beam-delivery system & 40.44 & 5,196 & 6,406 & 7,707 &ILC  & Doubled for dual IP \cr 
    Post-BDS beamline & 40.44 & 346 & 427 & 514 & ILC  & Costed as BDS. \cr 
    Turn-arounds & 40.44 & 213 & 213 & 213 &  ILC & Costed as BDS \cr 
    Instrumented beamline &  15.40  & 437 & 666 & 966 &ILC & In between acc.~structures.\cr
    Transfer line & 15.40 & 6,087 & 7,294 & 8,732 & ILC & Costed as instrum.~beamline. Driver and $e^+$ transfer lines. \cr
    Plasma cells & 46.20 & 375 & 570 & 825 &  & 3$\times$ instrumented beamline \cr
    Interstage optics & 40.44 & 738 & 910 & 1095 & & Costed as BDS \cr
    Driver-distribution system (both sides of plasma linac) & 40.44 & 2,226 & 2,960 & 3,840 & & Costed as BDS. One on each side of the plasma linac. \cr
    Tunnel (4.0~m inner diam.) & 11.89  & 2,713 & 2,713 & 2,713 & CLIC & Outer diameter 5.1 m. Surface-to-underground and turnaround. \cr
    Tunnel (5.6~m inner diam.) & 20.19 & 560 & 560 & 560 & CLIC & Outer diameter 6.7 m. Damping ring and $e^+$ source and injector. \cr
    Tunnel (8.0~m inner diam.) & 37.15  & 4,951 & 6,525 & 8,403 & CLIC & Outer diameter 9.1 m. $e^-$ injector, plasma linac, $e^+$ RF linac, BDS. \cr
    Surface building &  33.26  & 1,267 & 1,944 & 2,830 & CLIC & Used for drive-beam linac \cr
    Cut-and-cover tunnel & 9.86  & 2,035 & 2,712 & 3,597 & CLIC & Used for drive-beam linac and combiner rings \cr
     \cr
    \hline
    
    {\bf Cost element (per volume)}&Cost/volume & \multicolumn{3}{c}{Volume (m$^3$)}  & Ref. & Comment \cr
    & (kILCU/m$^3$) & 250 GeV  & 380 GeV  & 550 GeV & &  \cr
    \hline
    Tunnel (boring machine) & 0.573 & 397,190 & 499,546 & 621,641 & CLIC & Based on outer diameter. \cr
    Tunnel widening (excavation) & 0.45 & 148,699 & 183,328 & 220,556 & FCC & Used in dual BDS widening. \cr
    Cut-and-cover tunnel & 0.45 & 44,589 & 59,423 & 78,814 &  & Estimate based on tunnel area. \cr
    \cr
    \hline
    {\bf Cost element (per power)} & Cost/power & \multicolumn{3}{c}{Power (MW)}  & Ref. & Comment \cr
    & (MILCU/MW) & 250 GeV  & 380 GeV  & 550 GeV & &  \cr
    \hline
    Main beam dumps & 2.39 & 12.8 & 19.5 & 28.2 &ILC &  \cr
    Driver dumps & 2.39 & 4.8 & 7.3 & 10.6 & & Based on main beam dumps \cr
    LN2 re-liquification plant & 13.5 & 2.5 & 3.9 & 5.8 & C$^3$ & Per power at cryo temp.~($\sim$15\% cooling eff.~at 77~K) \cr
    Klystron (S-band) & 0.009 & 20,787 & 31,173 & 44,775 & C$^3$ & Peak power \cr
    Modulator (S-band) & 0.006  & 20,787  & 31,173 & 44,775 & C$^3$ & Peak power \cr
    Klystron (L-band) & 0.015  & 8,528 & 13,137 & 19,165 &  CLIC & Peak power \cr
    Modulator (L-band) & 3.9 & 42.8 & 66.0 & 96.3 &  CLIC & Average power \cr
    \cr
    \hline
    
    {\bf Cost element (individual)}&Cost & \multicolumn{3}{c}{Power (MW)}  & Ref. & Comment \cr
    & (MILCU) & 250 GeV  & 380 GeV  & 550 GeV & &  \cr
    \hline
    Klystron (S-band, injectors) & 0.351 & 21 & 21 & 21 & C$^3$  & 39 MW peak, 28 kW avg. \cr
    Modulator (S-band, injectors) & 0.234 & 21 & 21 & 21 & C$^3$ & 39 MW peak, 28 kW avg. \cr
    Klystron (S-band, main linac) & 0.603 & 298 & 453 & 656 & C$^3$ & 67 MW peak, 38 kW avg. \cr
    Modulator (S-band, main linac) & 0.402 & 298 & 453 & 656 & C$^3$ & 67 MW peak, 38 kW avg. \cr
    Klystron (L-band, driver linac) & 0.409 & 409 & 630 & 919 & CLIC & 21 MW peak, 105 kW avg. \cr
    Modulator (L-band, driver linac) & 0.313 & 409 & 630 & 919 & CLIC & 21 MW peak, 105 kW avg. \cr
    Waveguides & 0.0273 & 728 & 1,104 & 1,596 & CLIC & Assumed same for L- \& S-band \cr
    Low-level RF components & 0.0455 & 728 & 1,104 & 1,596 & CLIC & Assumed same for L- \& S-band\cr
    Combiner ring RF kickers & 1 & 6 & 6 & 6 &  & Rough estimate (no source). \cr
    Polarized positron source & 178 & 1 & 1 & 1 & ILC & Helical undulator and target. ILC cost minus the RF injector. \cr
    Polarized electron source & 82 & 1 & 1 & 1 & ILC & Photocathode gun. ILC cost minus the RF injector. \cr
    Driver source  & 10 & 1 & 1 & 1 & & Thermionic gun with relaxed performance. Rough estimate only without source. \cr
    Dual IP interaction area  & 154 & 1 & 1 & 1 & CLIC & \cr
    Experimental area & 20 & 1 & 1 & 1 & CLIC & \cr
    \hline
    \end{tabular}
   \caption{Cost basis for the estimation of HALHF construction costs, based on CLIC \cite{CLIC_CDR_2012,CLIC_PIP_2018}, ILC \cite{ILC_2013}, FCC \cite{FCC_2019} and C$^3$ \cite{CCC_2023}.}
      \label{tab:cost}
\end{table}

\subsection{Full programme cost}

Additional considerations are required when attempting to locate the overall optimum collider parameters; if only the construction cost was used, a machine operating with low luminosity for, say, 1,000 years would appear optimal. Therefore, when performing the Bayesian optimisation, a more complete cost must be used. The merit function to be minimised has been defined as a "Full Programme Cost" (not unlike the US ``Total Project Cost") to deliver a physics programme corresponding to collecting 2~ab$^{-1}$ of data at 250~GeV or 4~ab$^{-1}$ at 550~GeV. This function is given by
\begin{align*}
    \mathrm{Full~Programme~Cost}=&\mathrm{~} \mathrm{Construction~Cost~(components~and~civil~engineering)} \\&\mathrm{~}+ \mathrm{Overheads~(design,~development,~management,~inspection,~etc.)} \\&\mathrm{~}+ \mathrm{Integrated~Energy~Cost~(until~integrated~luminosity~reached)} \\&\mathrm{~}+ \mathrm{Maintenance~Cost~(over~programme~duration)} \\&\mathrm{~}+ \mathrm{Carbon~Tax~(construction~and~operations~emissions)}
\end{align*}
The construction costs include all the machine components, the civil engineering (tunnels, surface buildings and the interaction region), as well as additional infrastructure and services. The latter consists of eight parts, seven of which are costed as a fraction of the total civil engineering cost (based on the CLIC recosting 2025 submitted to this process~\cite{CLIC2025}): electrical distribution (20.3\%), survey and alignment (15.7\%), transport and installation (4.8\%), safety systems (11.7\%), machine control infrastructure (1.2\%), and access safety and control systems (1.8\%). Additionally, the eighth part is cooling and ventilation, which scales with the integrated collider power (costed at 2.85~MILCU/MW). The additional overheads are estimated at around $\sim$22\% of the total construction costs (10\% for design/development, 12\% for management/inspection). For HALHF, this is around 696 / 900 / 1161~MILCU for the 250 / 380 / 550~GeV options.

Next, the integrated energy costs are given by
\begin{align*}
\mathrm{Integrated~Energy~Cost}=\mathrm{Collider~Power} \times\frac{\mathrm{Integrated~Luminosity}}{\mathrm{Instantaneous~Luminosity}}\times\mathrm{Energy~Cost~Rate} ,
\end{align*}
i.e., the collider power over the integrated uptime (running time needed to collect the required amount of data times the energy cost rate (costed at 50~MILCU/TWh; approximately that used by CERN). A preliminary estimate for HALHF, which would run for about 9--10~years, is 320 / 440 / 600~MILCU for the 250 / 380 / 550~GeV options.

The maintenance cost, which must be included to give appropriate weight to the overall runtime of the programme, thereby encouraging high instantaneous luminosity, is given by
\begin{align*}
\mathrm{Maintenance~Cost}=\mathrm{Personnel~(FTEs)}\times\mathrm{Labor~Cost} \times\frac{\mathrm{Integrated~Luminosity}}{\mathrm{Inst.~Luminosity}\times\mathrm{Uptime~Fraction}},
\end{align*}
where the personnel requirement is estimated to be 100~FTEs per BILCU of construction cost (costed at 0.07~MILCU/FTE-year), equivalent to 0.7\% of the construction cost per year (based on ILC), and the uptime fraction is assumed to be 70\%. For HALHF, this gives a preliminary estimate for an integrated maintenance cost (over $\sim$10 years) of about 217 / 266 / 332~MILCU for the 250 / 380 / 550~GeV options, respectively.

Finally, a carbon tax is included to give weight to carbon emissions (encouraging a ``greener" machine in the optimisation). The CO$_2$ emissions are dominated by electricity production and tunnel construction (mainly the concrete), as outlined by Breidenbach \textit{et al.}~\cite{Breidenbach2023} (the accelerator components themselves are currently not included). The carbon tax is therefore given by
\begin{align*}
\mathrm{Carbon~Tax}=&\mathrm{~}(\mathrm{Tunnel~Length}\times\mathrm{Construction~Emissions} \\& +  \mathrm{~Integrated~Energy~Usage}\times\mathrm{Electricity~Emissions} )\times\mathrm{Carbon~Tax~Rate},
\end{align*}
where we are using 6.38~kton~CO$_2$-equivalent (CO$_2$e) per km of tunnel based on CLIC (with only marginally larger emissions at 7.34~kton~CO$_2$e for wider tunnels as in ILC) plus another 73\% for additional structures: 11.0~kton~CO$_2$e per km of tunnel. The emissions from electricity is estimated to be 20~ton~CO$_2$e/GWh The carbon tax rate is estimated at 800~ILCU/ton (based on the European Investment Bank's estimate for 2050~\cite{EIB2050}). For HALHF, this amounts to a total carbon tax of about 175 / 227 / 295~MILCU for the 250 / 380 / 550~GeV options (preliminary estimate only).

In summary, our preliminary estimate for the Full Programme Cost for HALHF (including 10 years of operation) is 45\% higher than the construction cost, at around 4,570 / 5,923 / 7,663~MILCU for the 250 / 380 / 550~GeV options.

\section{Civil engineering design study at CERN}
\label{sec:civil}

\begin{figure}[!h]
\centering
    \includegraphics[width=\linewidth]{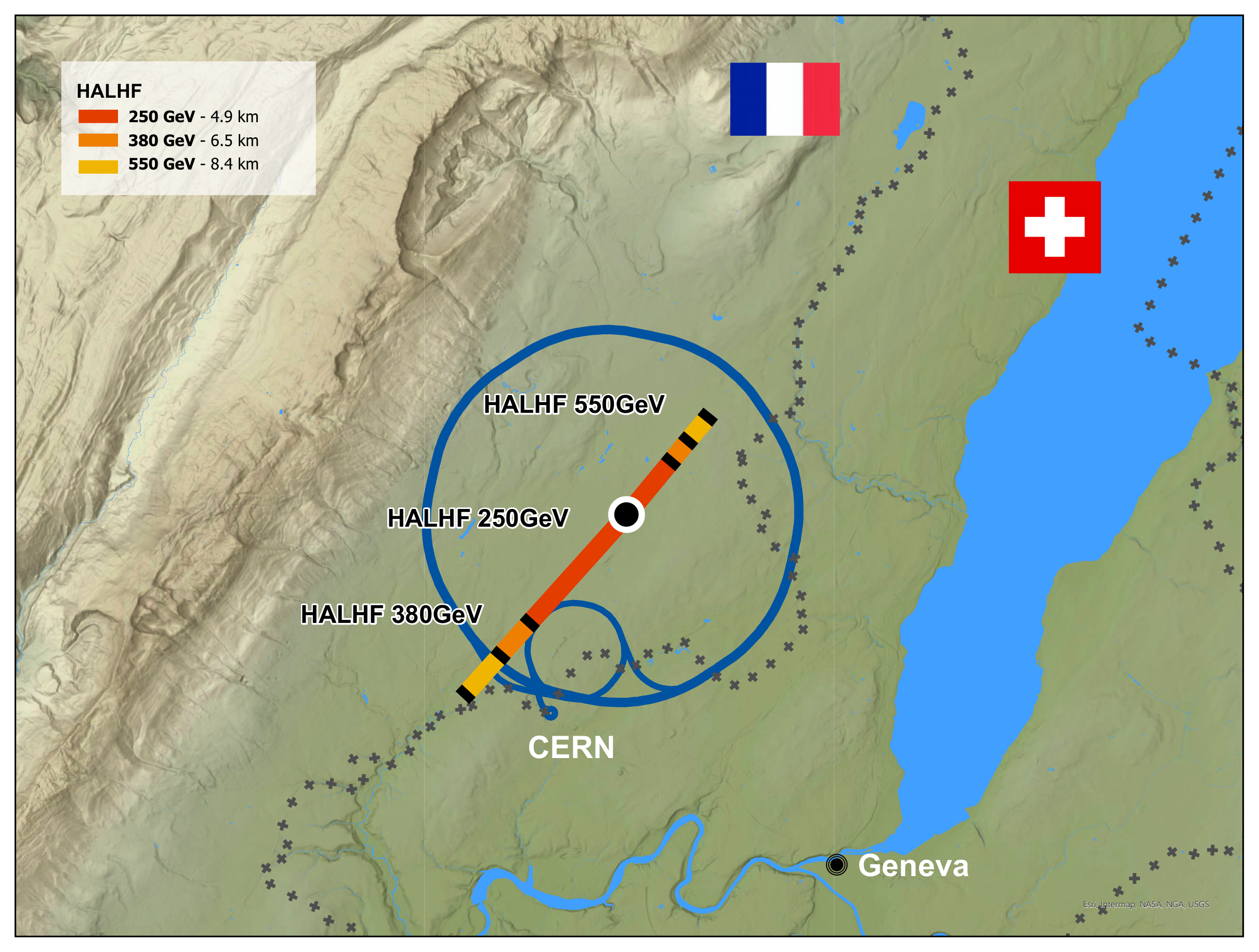}
    \caption{A schematic of the HALHF site studied at CERN (red, orange and yellow). Existing CERN infrastructure including the LHC is shown in blue.}
    \label{fig:footprint}   
\end{figure}

Civil Engineering (CE) represents a significant proportion of the implementation budget for tunnel projects such as the HALHF at CERN. As a result, CE studies are of critical importance to ensure a viable and cost-efficient conceptual design from the beginning. 
The baseline for the HALHF implementation at CERN is displayed in Fig.~\ref{Fig-1}. Initial placement studies at CERN have been conducted to find a suitable location to host the HALHF facility. The key driver in these studies was to align HALHF with the current ongoing CERN studies of CLIC \cite{CLIC_CDR_2012,CLIC_PIP_2018} and the LCF, most notably, sharing geological alignments and interaction regions.
Figure~\ref{fig:footprint} displays the proposed asymmetric HALHF alignment at its three energy stages. An important point to note is that the full-length machine would remain entirely within France with the tunnel situated in good ``molasse'' rock, well suited for tunnel-boring-machine (TBM) tunneling.

\begin{figure}[t]
\centering
    \includegraphics[width=0.93\textwidth]{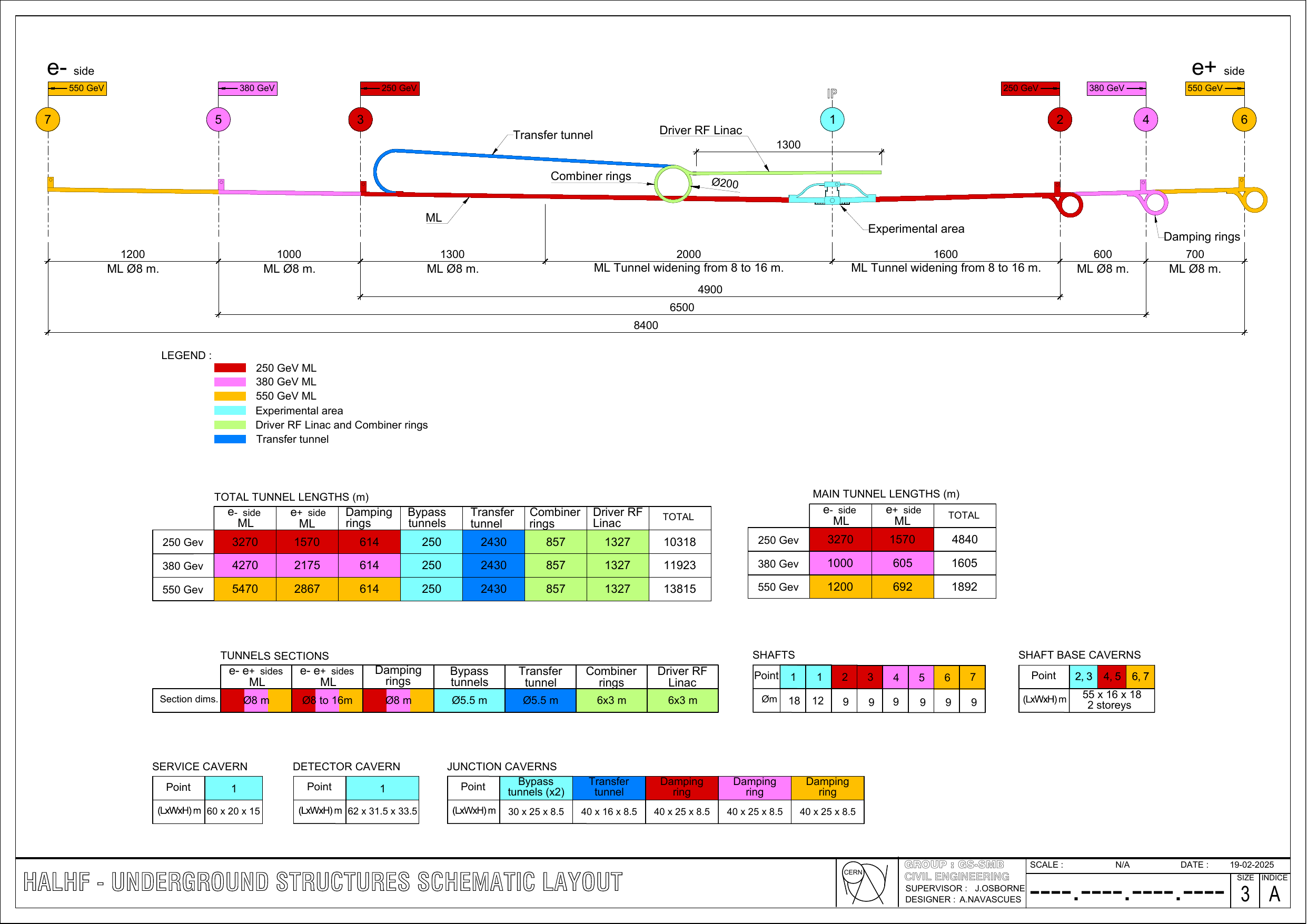}
    \caption{HALHF Underground Structures Schematic.}
    \label{fig:civil-schematic}   
\end{figure}

An underground structures schematic has been assembled in Fig.~\ref{fig:civil-schematic}. This displays the parameters associated with the different HALHF components and the different energy stages that follow, colour coded as shown. A main tunnel section of 8~m has been used (wider than CLIC's 5.6~m to allow for a PWFA driver-distribution system in the electron arm as well as fitting klystrons and modulators in the positron arm), which in the beam-delivery system (BDS) widens linearly to 16~m into the Interaction Region to accommodate the incoming beamlines to the two detectors. The Interaction Region follows that of CLIC, which has been updated in 2024 to house two offset detectors within a single cavern. 

\begin{figure}[!ht]
    \centering
    \includegraphics[width=0.93\textwidth]{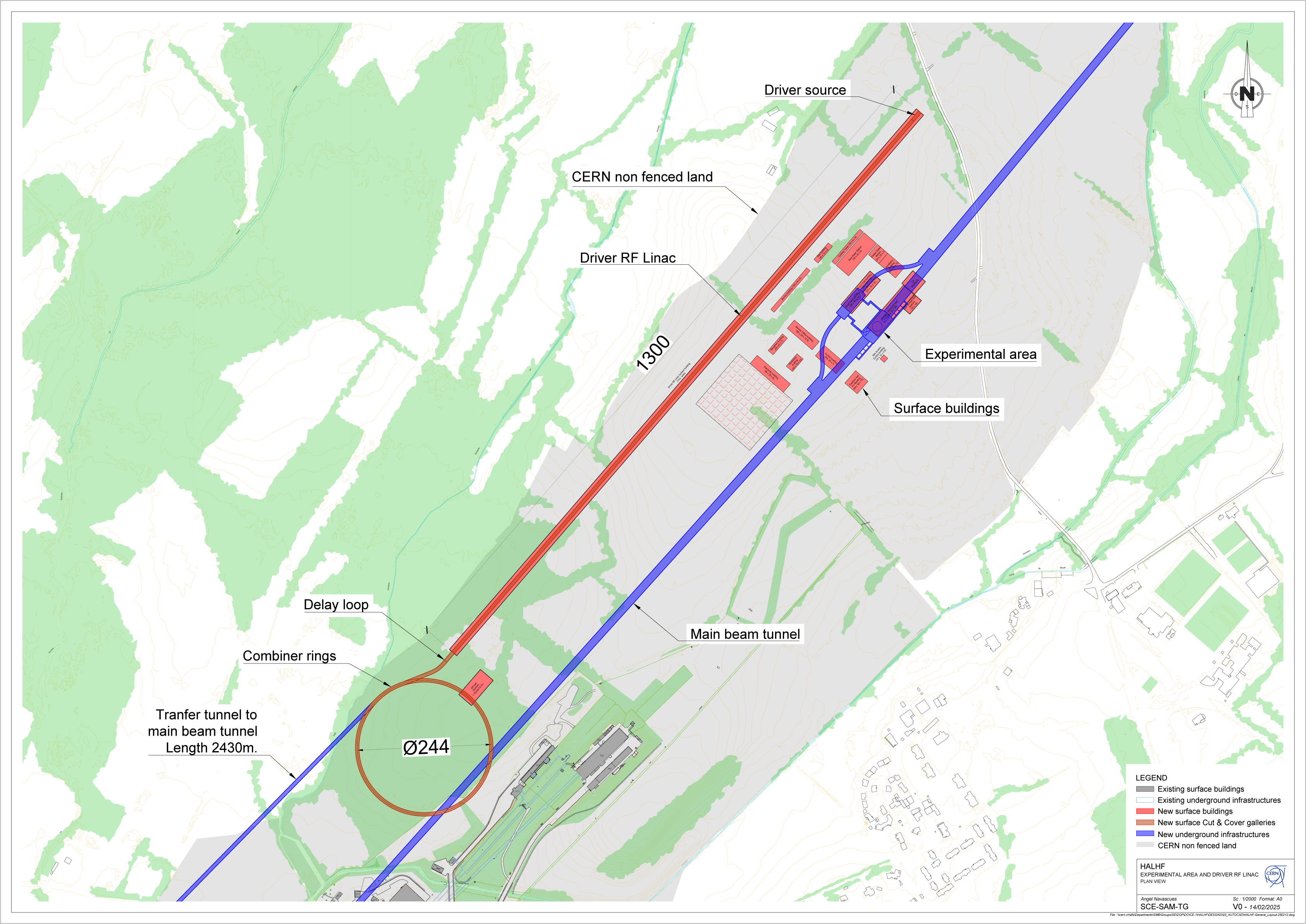}
    \caption{HALHF Injection Complex on the CERN Pr{\'e}vessin Site.}
    \label{fig:dbcomplex}   
\end{figure}

We have concluded that the Injection Complex (driver linac, delay loop and combiner rings) could be housed on existing CERN-owned land on the Pr{\'e}vessin campus in cut-and-cover tunnels. Figure~\ref{fig:dbcomplex} presents the Injection Complex alongside the Surface Experimental Area which is directly above the Interaction Region. The location of the Interaction Region is consistent across all linear collider studies at CERN, ensuring the Surface Experimental Area is located on land owned by CERN.

Beyond the Combiner Ring, a transfer tunnel connects the surface complex to the accelerator tunnel approximately 125~m below the surface. The full layout of the 250~GeV stage is portrayed below in Fig.~\ref{fig:250GeVlayout}. The transfer from the surface injects directly into the plasma-accelerator linac. Two shafts are required at the Interaction Region, an experimental shaft for the installation of the detectors, and a service shaft to provides access into the region. Alongside these shafts, two further shafts are required at the extremities of the tunnel for servicing access and safety.
 
The CLIC shafts are separated by up to 5.7~km; thus, if the HALHF 550~GeV stage was to be built first, only four shafts in total would be required for the whole project. If the energy stages where to be constructed sequentially, eight shafts would be required to reach the 550~GeV stage. 
This principle also applies to the transfer tunnel from the surface to the beginning of the plasma-accelerator linac. If the civil engineering works for HALHF at 550~GeV could be completed in one step, a single transfer tunnel would be constructed for the whole project. If the three stages were constructed sequentially, three transfer tunnels would need to be constructed.
In summary, building the full CE works to 550~GeV in the first stage would require a greater initial capital investment but would yield long-term savings, particularly by reducing the required number of shafts and transfer tunnels.

\begin{figure}[!ht]
    \centering
    \includegraphics[width=\textwidth]{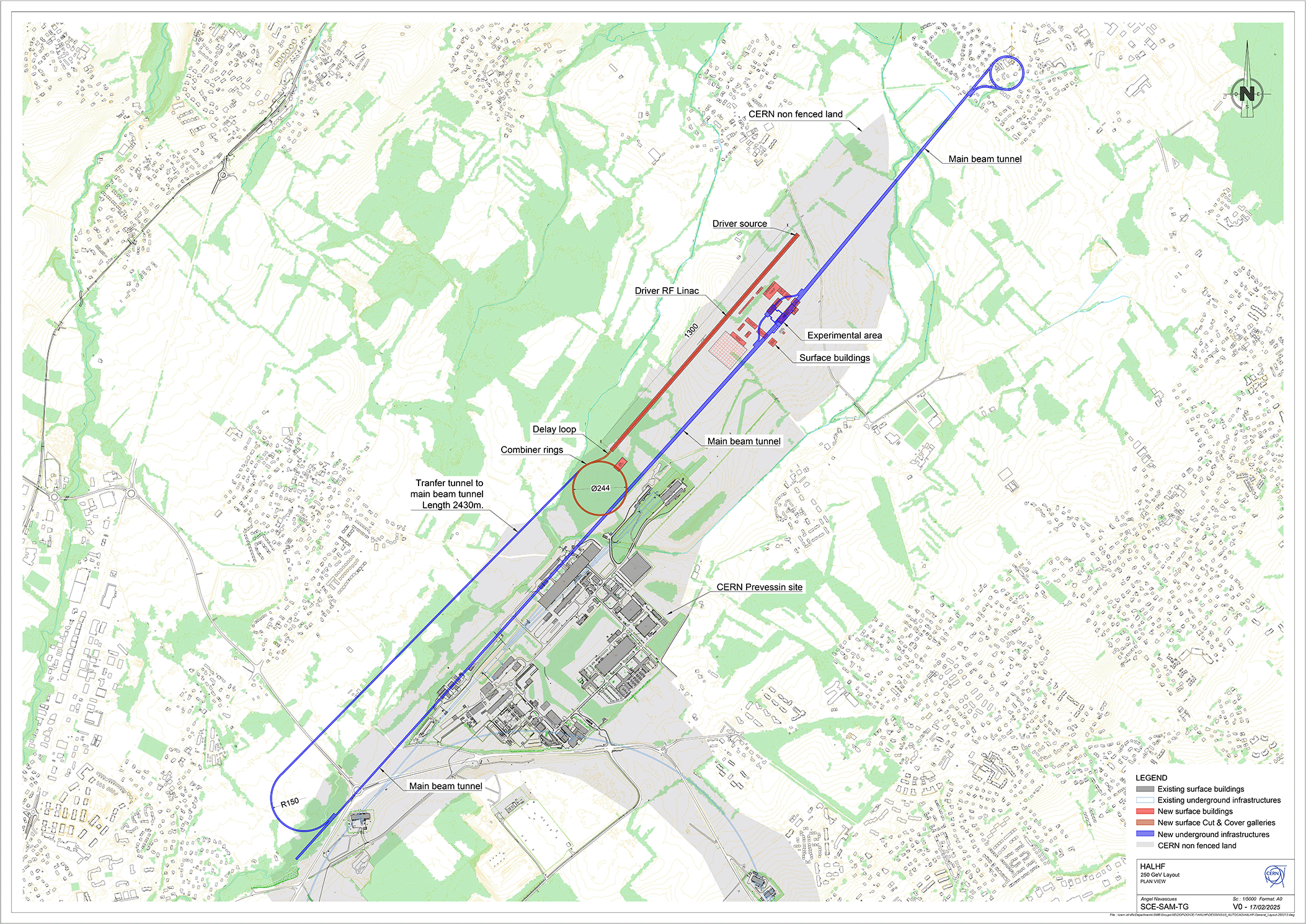}
    \caption{HALHF 250~GeV Machine Layout.}
    \label{fig:250GeVlayout}   
\end{figure}  

\section{Detailed description of subsystems}

This section describes in greater detail each subsystem of HALHF. It is based largely on the recently published Ref.~\cite{Foster2025}, but adds new information on some parts, including the BDS.

\subsection{Electron sources}
\label{sec:electronsource}
Two electron sources are required for HALHF. The first is the source that produces a high-degree-of-polarization and low-emittance bunch that is accelerated in the PWFA arm and collided with a positron bunch at the IP. Such a high-quality polarized source, which produces an emittance small enough that a damping ring is unnecessary, does not yet exist, although unpolarized sources with such performance do~\cite{Xu_DR_Free}. The main problem with polarized sources is reduced lifetime due to eventual poisoning of the photocathode. Development of the required polarized sources is a subject of considerable R\&D~\cite{Maxon2024} and we assume that on the timescale of HALHF the required sources can be developed for other applications. If not, the addition of an electron damping ring is a relatively minor perturbation both on layout and cost.

The electron source for the drive beam does not require either polarization or low emittance. A standard high-current thermionic source is therefore proposed, very similar to that used by CLIC for drive-beam generation~\cite{CLIC_PIP_2018}.

\subsection{Positron source}
\label{sec:positronsource}

\begin{figure}[ht]
\centering
\includegraphics[width=\linewidth]{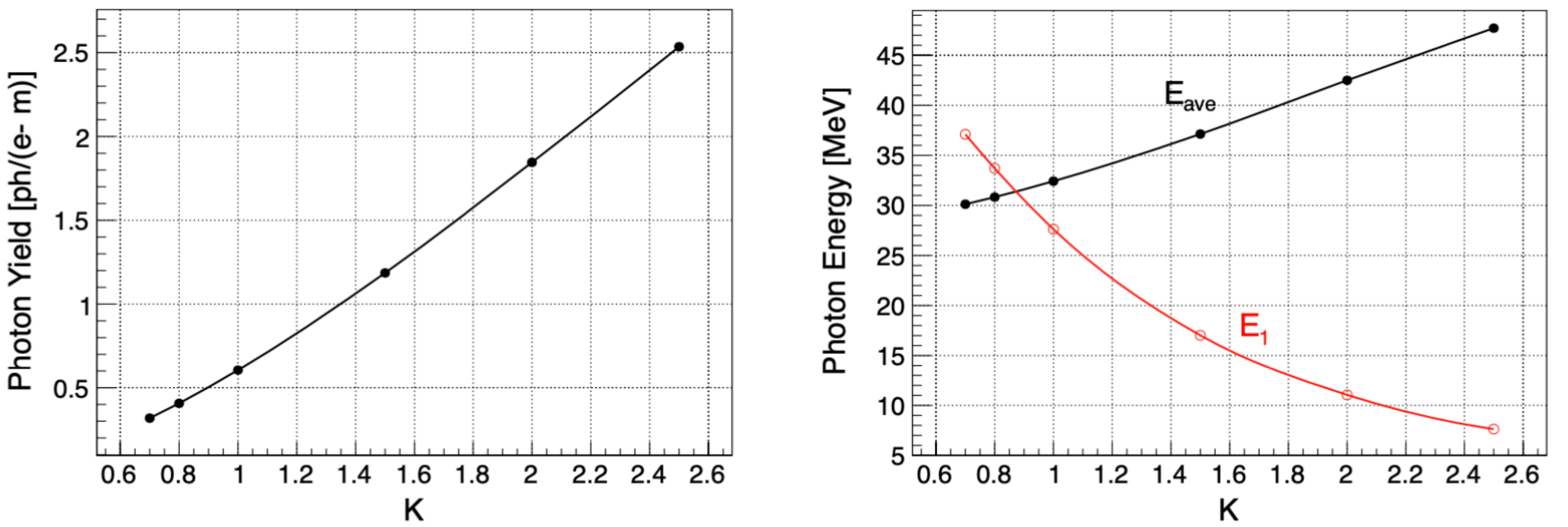}
\caption{Photon yield (left) and photon energy (right) as a function of
the undulator $\kappa$-value. $E_{\mathrm{ave}}$ is the average photon energy and $E_1$ 
is the energy cut-off of the 1st harmonic. From Ref.~\cite{Ushakov:2013bm}.}
\label{fig:6_undulator}       
\end{figure}

\begin{figure}
\centering
\includegraphics[width=0.8\linewidth]{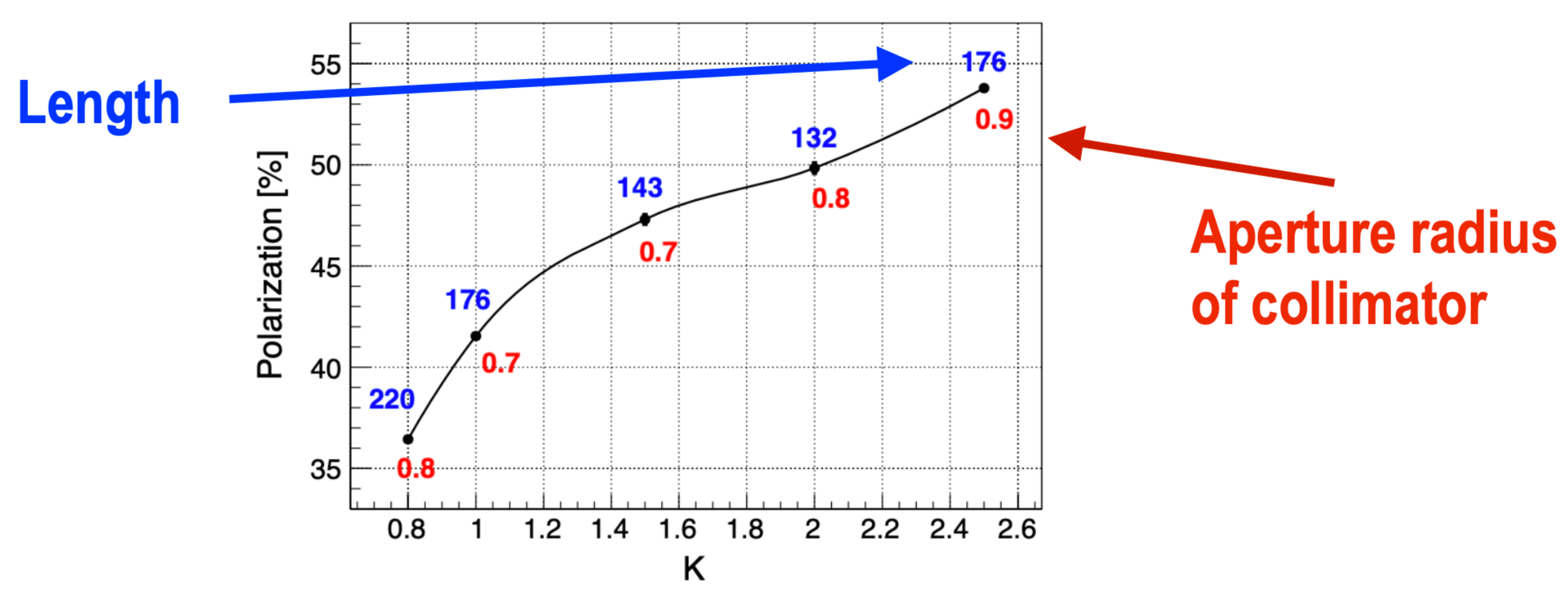}
\caption{Positron polarization versus $\kappa$-value without collimation but with an optics-matching device generating a peak field on the beam axis of $\sim$3.2~T. The blue numbers indicate the required undulator length. From Ref.~\cite{Ushakov:2013bm}.}
\label{fig:7_polarization}      
\end{figure}

The positron source for HALHF, and indeed for any \ee\ linear collider, is challenging. This is particularly the case when polarized positrons are an important factor in physics reach~\cite{Moortgat-Pick:2015lbx}. Fortunately, a source with similar characteristics to that required for HALHF has already been substantially designed for ILC. 

In the ILC design, a 125~GeV ${\rm{e}^-}$-beam passes through a superconducting helical undulator, generating circularly polarized photons with energy $\sim$7.5~MeV that impinge on a thin rapidly rotating target constructed from a Ti alloy, producing polarized electrons and positrons. The positrons are captured, pre-accelerated and led through spin rotators before entering damping rings. Advanced simulations of the undulator have been performed~\cite{Alharbi:2024dud}, e.g.~examining the impact of field misalignments, errors in the magnetic field ($\kappa$-value), and the period $\lambda$. Depending on the $\kappa$-value, such uncertainties can affect the attainable polarization and the load on the target.

The following subsections describe work carried out to strengthen the ILC design and then its adaption for HALHF. 

\subsubsection{Rotating target}
\label{sec:rotatingtarget}
The 1~m-diameter target for ILC was foreseen to rotate on magnetic bearings at 2000~rpm, which corresponds to a tangential speed of 100~m/s. This leads to the photon beam returning to the same target position every 7 seconds. The beam has a power of $\sim$60~kW, but only about 3\% of this is deposited in the target. Radiative cooling is sufficient within the vacuum chamber~\cite{Riemann:2020ytg}; the heat from the vacuum chamber is taken away by water cooling. Discussions on manufacturing the device are ongoing with the SKF company in Canada. 
 
\subsubsection{Mask design} 
\label{sec:maskdesign}
A mask system to protect the undulator walls has been studied, designed to restrict  the synchrotron radiation deposition to <1~W/m, even at the maximal length of 320~m for the ILC. 

\subsubsection{Optics-matching device}
\label{sec:opticsmatch}
There has recently been substantial progress towards manufacturing a scaled-down prototype pulsed solenoid used as an optics-matching device, suitable for 1~ms photon pulses hitting the target. Manufacturing drawings have been produced and prototypes manufactured using 3D-printing. First measurements of the fields with 1~kA (pulsed and DC) are planned in 2025, which will be extended to 50~kA at CERN. The higher yield required for HALHF (estimated to 3--4 positrons/electron), greater than the 1.5~positrons/electron required for ILC, should be achievable and is currently being studied~\cite{Moortgat-Pick:2024fcy}. 

\subsubsection{Undulator-based positron source for HALHF}
\label{sec:undulator-HALHF}
The fully accelerated HALHF $e^-$ beam has an energy of 375~GeV. This would be used as the photon source in the undulator; this has much higher energy than the ILC Higgs-factory design, so that the undulator parameters require adjustment.

The parameters calculated for the original 1~TeV ILC upgrade option~\cite{Ushakov:2013bm} can be used as a starting point, i.e.~an undulator with 174~m length, a period of $\lambda=43$~mm and a high magnetic field with $\kappa=2.5$.
The undulator radiation is simulated using Kincaid's formulae~\cite{Kincaid:1977fg}.
The photon-generation efficiency in such an undulator as a function of 
$\kappa$ is shown in the left-hand plot of Fig.~\ref{fig:6_undulator}, where the photon yield has been normalised per electron per meter of undulator. The photon energy cut-off of the 1st harmonic and the average photon energy are shown in the right-hand plot. The impact of the undulator field on the $\rm{e}^+$ polarization is shown in Fig.~\ref{fig:7_polarization}.

A parameter set such as that for ILC at 1~TeV could produce polarization of up to 54\%.  However, the high $\kappa$ value means that higher harmonics are important, leading to higher mean power and greater energy spread for the photon beam. This makes the $e^+$ capture more difficult but the use of e.g.~a pulsed solenoid or indeed a plasma lens give grounds for optimism. 

Detailed simulations using \textit{CAIN} adapted for undulator radiation~\cite{Yokoya} are currently being carried out using the above scheme for HALHF. It is expected that estimates for the achievable HALHF positron yield will soon be available. The requirement is to produce in excess of three photons per electron, which will certainly be challenging. 

\subsection{Drive-beam linac and combiner rings}
\label{sec:drivebeamlinac}

The parameter set from the Bayesian optimiser described in Sect.~\ref{sec:parameteroptimisation} produced parameters for the drive-beam linac and combiner rings that are very similar to those produced for the CLIC drive beams. This is reassuring, since the latter have been the subject of many person-years of design work and the purpose, to transfer power from low-energy high-charge beams to high-energy low-charge beams to be collided, is very similar. 

\begin{figure}[h]
    \centering
    \includegraphics[width=\textwidth]{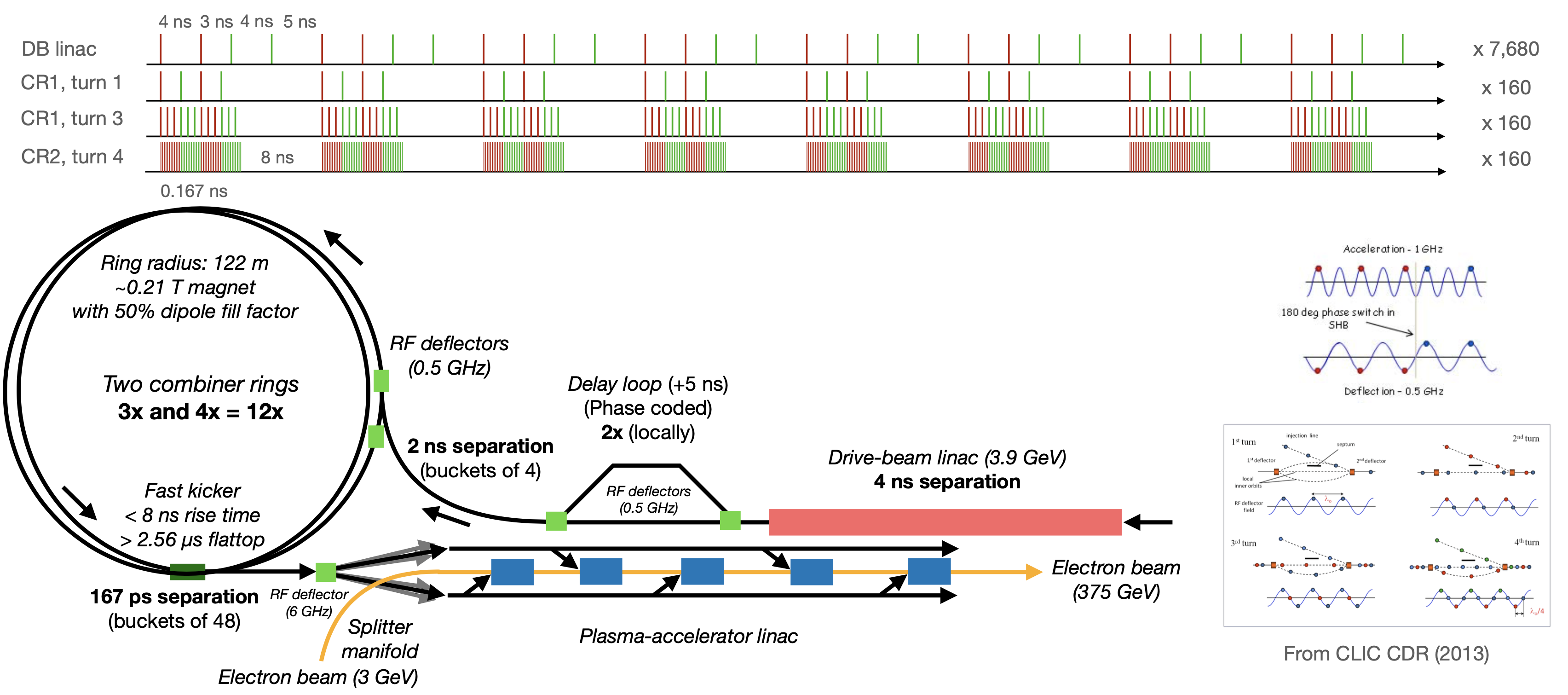}
    \caption{The drive-bunch structure for HALHF. Bunches generated with an average separation of 4~nsecs are interleaved in a delay loop and two combiner rings to give bunch trains with 48 bunches separated by 167~psecs. The bunch trains are then extracted and sent via the distribution system to each PWFA cell with the appropriate timing. The two insets on the right illustrate the combining function for CLIC, which is identical to that for HALHF.}
    \label{Fig-db_bunch_structure}       
\end{figure}

There are however some differences from the CLIC drive-beam parameters. The drive beams for HALHF are significantly more energetic, 4~GeV rather than 2~GeV. This has consequences in that a higher gradient of 4~MV/m is required and the combiner-ring radius is larger. Otherwise the pattern of combining pulses necessary to reduce the peak load on the linac is identical to that of CLIC. The HALHF bunch structure is also different, with bunch trains of 48 bunches separated by 167~ps. The bunch structure is shown schematically in Fig.~\ref{Fig-db_bunch_structure}.

\subsection{Drive-beam distribution}
\label{sec:drivebeamdist}
Distributing drivers to the plasma stages could be based on a previously presented scheme~\cite{Adli2013}. This involves a periodic, undulating delay chicane with fast kickers (1--4~ns rise time) to extract the rear-most driver into each stage. The proposed setup has two such chicanes: one on each side (left and right) of the plasma accelerator, as this produces an on-average straight multistage linac even when the interstage optics has an angle. Bending the drivers enough to delay 1--2~ns per stage is facilitated by reducing the driver energy as much as possible; it should not be greater than around 10~GeV.  

The use of fast kickers in the distribution system may not be ideal. Given that trains of drivers need to be accelerated in a continuous train without gaps, as this is the most efficient way to produce drivers in an RF linac, the kickers need to have rise and fall times on the ns scale and repeat such kicks hundreds of times with a period that is sub-100~ns. While this is not currently easy to do with strip-line kickers, it may be possible with harmonic RF kickers~\cite{Huang2016}. However, having many such complex devices (one per stage) could be troublesome. The stability of these kickers will also need to be very high, as this directly influences the orbit of the driver in the plasma stage. Overall, while it may be possible to use an undulating chicane with fast kickers, other solutions should be explored.

One possible alternative solution separates the drivers into parallel beam lines in a single multi-beam ``splitter” just prior to the plasma stages, requiring more transport lines and delay chicanes, but no fast kickers. It is unclear whether the delay chicanes should bend the drivers in the horizontal or vertical plane. Arguments for bending in the horizontal plane include that this improves the orbit stability in the vertical plane. This is important for the plasma stages since the vertical emittance has much stricter tolerances. On the other hand, the driver should have a lower emittance in the horizontal plane to avoid resonant emittance mixing~\cite{Diederichs2024}, which is easier to maintain when bending in the vertical. In addition, vertical bending would require less horizontal space, which is always at a premium in the beam tunnel. 

Deflectors running at 1~GHz are a possible solution. Depending on the phase of the deflector, bunches are steered alternately to each side of the central axis along which the electron beam to be accelerated and collided passes. Further deflectors of the appropriate frequency then steer the drive-beams to a dipole that deflects them to the correct plasma cell with the appropriate timing to generate the accelerating bubble for the colliding electron bunch. The necessary timing accuracy is $\sim$10~fs. Such precision is achievable with state-of-the-art synchronisation; however, the detailed design and operation of this scheme remain to be finalised. 

\subsection{PWFA linac}
\label{sec:PWFA}

The plasma linac accelerating high-energy electrons for HALHF will require multiple
plasma-accelerator stages. This is a highly non-trivial aspect of the design that requires
re-thinking how to build a high-energy linac, including beam optics, diagnostics,
collimation, beam dumps etc. Fortunately, the plasma linac does not dominate
either the length or the cost, which gives some freedom in how it can be designed. Overall, it is clear that there is much “uncharted territory” and that a substantial dedicated R\&D programme will be
necessary.

\subsubsection{Staging optics and nonlinear plasma lenses}
\begin{figure}[b]
    \centering
    \includegraphics[width=\linewidth]{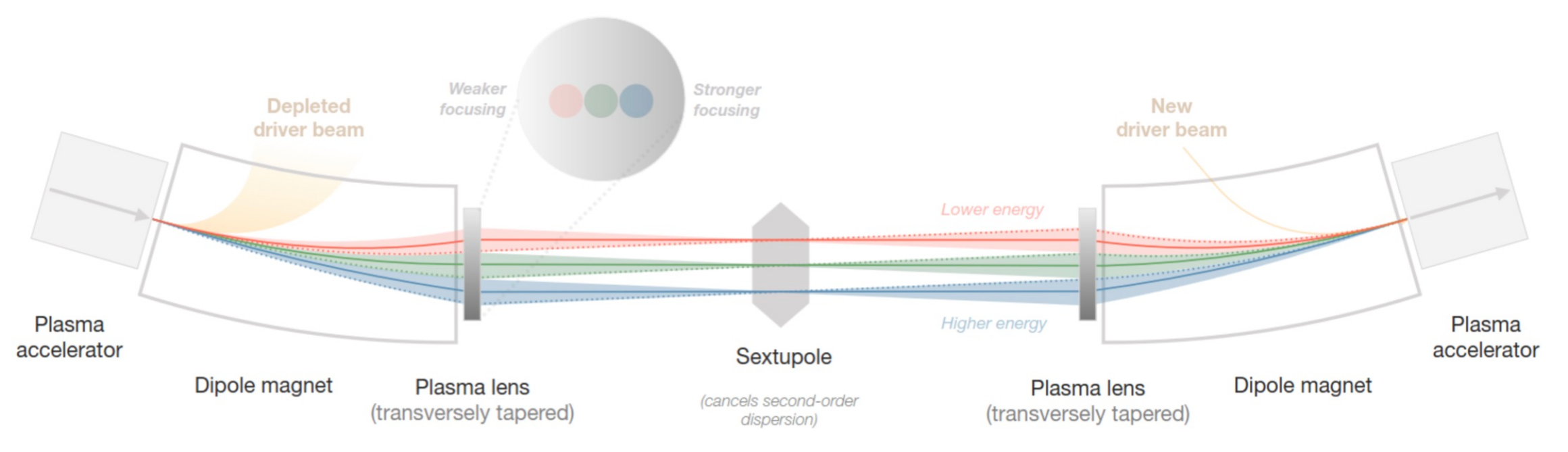}
    \caption{Schematic of an achromatic staging optic based on nonlinear plasma lenses. From Ref.~\cite{Drobniak2025}.}
    \label{fig:9_staging}      
\end{figure}

One of the key requirements of the plasma linac is to deliver stable, low-emittance electron beams, albeit with parameters somewhat relaxed compared to symmetric colliders, because of the gain in geometric emittance from working with higher-energy electron beams. While this reduces emittance-preservation requirements in or between every stage, there is nevertheless a strict emittance budget. Chromaticity is the key challenge for the interstage optics, due to the combination of divergence and energy spread in plasma accelerators. The planned solution is to use nonlinear plasma lenses~\cite{Drobniak2025}, which can in principle provide achromatic point-to-point imaging between the plasma-accelerator stages. One question is how the interstage optics transport the non-Gaussian transverse phase-space distribution that arises in the presence of ion motion; the interstage optics should be able to preserve it owing to the achromatic point-to-point imaging (see Fig.~\ref{fig:9_staging}).

The effect of plasma ramps needs to be assessed in this context, as they cause the beam to experience a time-varying ion density due to ion motion. Other important points include emittance growth from wakefield distortions as well as Coulomb scattering in plasma lenses; this must be studied in PIC simulations as well as codes such as \textit{GEANT4}~\cite{Allison2016} to model scattering accurately, as this may result in emittance growth, in particular because the beta function is large inside the plasma lenses. 
Beam–gas scattering inside plasma lenses could be important and requires a thorough investigation. Finally, the repetition-rate limitations of plasma lenses are not known and may restrict the allowable parameter space of bunch-train patterns in HALHF.

\subsubsection{Diagnostics between stages}
In order to preserve the emittance between stages and transfer beams into the next stage, sufficient diagnostics must be included. These include: an orbit measurement, using beam-position monitors (BPMs); an energy-spectrum measurement, likely using an insertable screen at the centre of the interstage optics (close to the central sextupole); and an emittance measurement, possibly using the same screen setup. We note that cavity BPMs can in principle measure angles as well as offsets, which could reduce the number of BPMs required. In the currently simulated beam optics, about 10\% of the space is left open for such diagnostics. 

There is also a requirement for an online/non-destructive measurement of the energy spectrum and emittance. This could potentially be achieved using a scintillator screen placed close to the beam, measuring its electric-field halo. Another possibility would be to use the betatron radiation exiting each stage for diagnostic purposes, as the spectral and spatial distribution of this radiation depend on beam parameters, such as orbit and
emittance.

One possibility that may ease commissioning is to have a dedicated diagnostics/tune-up station between some or all stages. Sending the beam to this station would require flipping the polarity of one of the interstage dipoles. The drivers also require diagnostics as they enter and exit the plasma stages. Beam-position monitors should be located before and after each stage; a natural location for a post-plasma spectrum measurement is the entrance of the driver-beam dump, onto which the energetically dispersed driver is anyway steered.

\subsubsection{Collimation}
\label{sec:collimation}
While collimation is not a requirement within the plasma linac, in practice multiple plasma accelerators with complex beam optics between them will present a narrower aperture than collimators in the beam-delivery system. This presents a new possibility: to do away with at least some of the (very lengthy) ``conventional” collimator system integrated into the BDS in favour of a distributed collimation system throughout the plasma linac. There are several benefits to such a setup: the  halo furthest distant from the beam core is removed first, at low energy, followed by a gradual removal of the remaining halo at higher energy. 

Two types of collimation will be required in the interstage: energy collimation and beta (transverse-phase-space) collimation. The energy collimation can be performed at the location of the central sextupole, as this location has large dispersion and small beam sizes—ideal for energy collimation. The beta collimation is likely to be done best at the location of the plasma lenses; these anyway impose a radius restriction on the beam. One remaining issue is that of collimating the transverse phase space at different phase advances: the plasma lenses are mainly located at the same phase advance (modulo 180$^\degree$), which makes it challenging to collimate the phase 90$^\degree$ away.

\subsubsection{Beam dumps, heat handling and the radiation environment}
The above discussion on collimation requires also a strategy to deal with the collimated particles. Unless these are directed into dedicated beam dumps, they will lead to heating of the surrounding area, but more importantly activate the plasma-accelerator components. 
The largest issue in this regard is however the dumping of the depleted drivers. Multiple megawatts of dumped beam power---most of it at sub-GeV energies—--will need to be handled. A large amount of cooling will also be required. Dedicated radiation simulations using \textit{FLUKA}~\cite{Ballarini2024} are planned to design the appropriate beam dumps.
\begin{figure}[!ht]
\centering
    \includegraphics[width=0.8\textwidth]{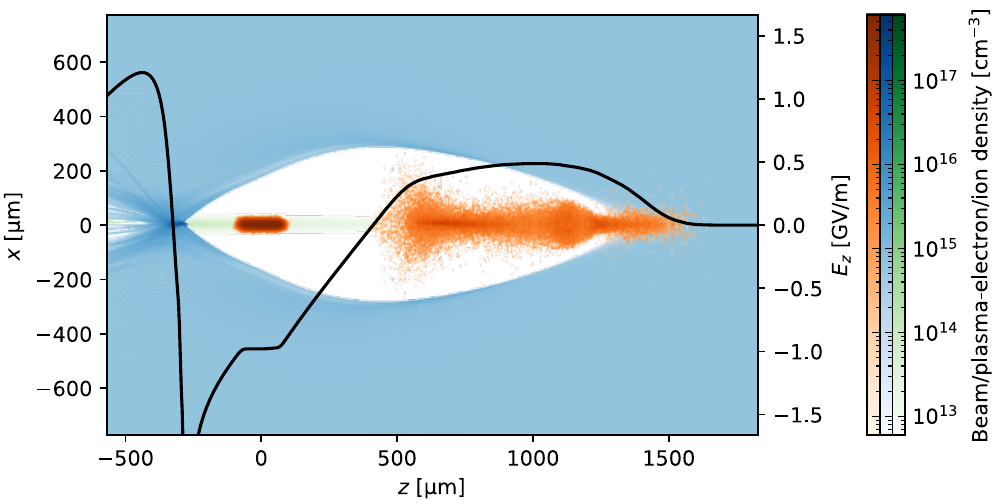}
    \caption{PIC simulation of the accelerating structure proposed for HALHF, with an 8/1.6~nC driver/trailing bunch (orange colour map) and a cold helium plasma (blue) of density \SI{6e14}{\per\cm\cubed}. A ramped driver current profile results in a flattened decelerating field and an 80\% energy-depletion efficiency. Ion motion (green) suppresses the beam-breakup instability.}
    \label{fig:pwfa_sim}   
\end{figure}
\subsubsection{Beam quality}
\label{sec:beamquality}
Several sources of emittance growth need to be considered for a PWFA collider \cite{LindstromThevenet_2022,Lindstrom2024}. Some are well known from RF-based linear-collider studies, including chromatic emittance growth and the wakefield effect due to beam misalignment. In the plasma bubble, the transverse fields, both focusing and the wakefield effects, are stronger than in an RF collider \cite{Lebedev2017,Finnerud2025}. The effect of these on the beam therefore needs special attention. For example, transverse wakefields are expected to be very strong, and betatron radiation is an effect that will be important in PWFA colliders, especially towards TeV-scale energies.  There are also additional effects on the beam in PWFA colliders related to ion motion and the plasma interstages.
 
Tools are being developed to simulate all the relevant effects. The PIC-code \textit{HiPACE++}~\cite{Diederichs2022}, seems to include the relevant physics for simulations of a single plasma stage in HALHF. One issue that requires further study is the necessity for symmetrical drive beams, which need to be guided to avoid transverse drifts due to misalignment. Multi-stage effects must  be studied along the whole linac, consisting of many stages and interstages. These studies are planned to be done with simplified plasma-stage models in the \textit{ABEL} start-to-end simulation framework, under development at the University of Oslo. One of the features of \textit{ABEL}, which is a multi-level framework that also integrates within it \textit{HiPACE++} and other codes, is a fast model including transverse wakefields and betatron radiation damping that has been benchmarked with \textit{HiPACE++}. An ion-motion model is currently under development.  The interstage part of the simulations is done in the conventional tracking code \textit{ELEGANT}~\cite{Borland2000}, with an additional element modelling an idealised nonlinear active plasma lens.
 
Preliminary results from \textit{ABEL} simulation scans indicate that the emittance growth due to transverse wakefields can be sufficiently mitigated with the help of the decoherence effect of ion motion. Studies are ongoing to quantify the tolerances related to wakefields, and also the effect of main-beam misalignment due to drive-beam jitter. The drive beam generates its own focusing axis, and even a very small jitter angle may lead to an unacceptable main-beam misalignment. This effect may be mitigated using magnetic guide fields around the plasma stage, and will be studied using \textit{ABEL}.
 
\subsubsection{Spin-polarization preservation}
\label{sec:polarization}
 
Although there is in principle no reason for plasma acceleration not to preserve polarization, this has to be demonstrated in simulations and in experiments.  Published results of simulations~\cite{Vieira2011} indicate already that preservation is possible at the required ILC levels ($>80$\%) at the output of the plasma linac, although the level of preservation depends on the emittance of the accelerated beam.
 
 The simulation work assumes P = 85\% at the start of the plasma linac. The permitted depolarization in each stage is then $(1-0.8/0.85)^{1/n}$, where $n$ is the number of stages, giving e.g.~1.2~permille relative decrease per stage over e.g.~48 stages.  In addition, interstage transport may contribute significantly through synchrotron radiation and nonlinear fields. This needs to be studied and made part of the polarization budget.
 
In an accelerator, the spin will precess according to the Thomas-BMT equations~\cite{Thomas1927, Bargmann1959}, which describe the precession of the spin of a charged particle in electric and magnetic fields. In a plasma bubble (i.e.~the blow-out regime, as used in HALHF) the equations may be greatly simplified, and the spin precession will depend only on the radial focusing force, i.e.~the off-axis position within the bubble. For flat beams ($\sigma_y \ll \sigma_x$), having polarization along the $y$-axis in the plasma seems ideal, as the precession of the $y$-component of spin is minimised. The BDS would then need to rotate the spin to give longitudinal polarization in the collisions.
 
Spin-transport (the simplified Thomas-BMT equations) will be implemented as simplified models in \textit{ABEL}. This will allow for start-to-end simulations, and the level of spin preservation in HALHF may be quantified. Spin-transport is already implemented in \textit{HiPACE++}.
 
While study with simulations is important, facilities/experiments to demonstrate experimentally the conservation of polarization are essential. A challenge is the limited number of polarized beams available for experiments world wide. In the short term, a relatively feasible experiment may be to test polarization preservation in an active plasma lens at the ELSA or MAMi accelerators at Bonn and Mainz, respectively. In the longer term, work on highly polarized plasma photocathodes is vital and will hopefully be pursued in several laboratories. An implementation of the SPARTA project \cite{SPARTA} with a polarized front-end would be the ultimate demonstrator for HALHF.

\subsection{Plasma generation, heating, cooling and power flow, efficiency}
\label{sec:heating}
Plasma accelerators can in principle achieve energy-transfer efficiencies comparable to traditional radio-frequency machines~\cite{Tzoufras2008}. However, due to the nature of the plasma-acceleration process, a certain proportion of the driving beam’s energy will remain in the plasma after the wakefield has passed. What happens to that energy, i.e.~how it is transported and on what timescale, will have implications for the choice of plasma-source technology and how that plasma source is operated. The following subsections examine these topics and their inter-relatedness in the context of HALHF.

\subsubsection{Efficiency}
The overall energy-transfer efficiency, i.e.~the energy gained by the accelerating bunch divided by the input driver energy, at HALHF is expected to be 40\%, with another 40\% of the initial drive-beam energy remaining in the plasma. Separate experiments have already shown that 57\% of the drive-beam energy can be transferred to the plasma~\cite{Pena2024} and that 42\% of the deposited drive-beam energy can be converted to gain in witness-beam energy~\cite{Lindstrom2021}. Combining these figures suggests an overall efficiency of 24\% is possible, so the HALHF proposal is ambitious but realistic. While a high efficiency of driver energy to the witness beam reduces the heating rate, moderate changes in its value will not meaningfully alter the scale of the technological challenge, which requires plasma devices capable of withstanding orders-of-magnitude higher heating loads than the current state of the art. Fortunately there is the potential to make rapid progress in this area with simple ideas, numerical simulations, and experimentation.

\subsubsection{Plasma generation}
Two common methods of field ionisation have been considered for HALHF: laser or high-voltage-discharge ionisation\footnote{AWAKE is investigating plasma generation with both discharge and Helicon sources in the range 1--\SI{10e14}{\per\cubic\cm} with a baseline of \SI{7e14}{\per\cubic\cm}~\cite{Buttenschon2018, Ariniello2019, Torrado2023}}. The target density envisaged for HALHF is \SI{\sim e15}{\per\cubic\cm}, which is compatible with both generation mechanisms. A pre-ionised plasma source, rather than relying on beam ionisation, is preferred to combat driver head erosion. \textit{Laser ionisation} is potentially the more flexible of the two options, as advanced shaping methods (with the use of Bessel beams for example) may give control over the shape of the plasma density ramps, which are crucial for emittance preservation. Additionally, plasma of any (sub-critical) density can be ionised. In order to ionise Ar or H, a focused intensity of around \SI{2e14}{W cm^{-2}} is required. The laser energy of a 40~fs (Ti:sapphire) laser required to ionise a 7.8~m long stage was calculated assuming a plasma column radius of twice the blowout radius (assumed to be 0.42~mm). A Bessel beam to do this would require 8 rings, each of energy 55~mJ. Therefore, a laser delivering >0.5~J (plus some conservative safety factor) would be required, which at 16~kHz (average) repetition rate is far beyond the state of the art. Furthermore, such lasers typically run with constant (CW) pulse spacing, perhaps excluding compatibility with the pulsed operation of the drive-beam and plasma linac. However, in the next section, it will be shown that the energy deposition from the train of driver bunches should keep the plasma hot enough to remain fully ionised, meaning that plasma confinement may be a more pressing issue than ultrahigh-repetition-rate plasma generation, and that only one external ionisation event may be required per bunch train. \textit{Discharge ionisation} is relatively simple, although the downside is that it favours higher densities (to stay close to the Paschen curve minimum in a 5 m plasma, a initial density of order \SI{e16}{\per\cubic\cm} is required). Strategies to mitigate this initial density requirement exist, such as using a hot gas, employing a glow discharge or localised laser ionisation to initiate the discharge. Scaled calculations from FLASHForward measurements~\cite{Loisch2025} indicate that the plasma required for HALHF can be produced with a few tens of Joules. The ease and low-energy requirements of discharge ionisation recommend it for the HALHF baseline design.

\subsubsection{Heating}
As mentioned in a previous subsection, 40\% of the drive-beam energy will be deposited in the plasma. This equates to $\sim$13~J of total energy per acceleration event (or $\sim$1.6~J/m averaged over the length of the plasma-acceleration module) reaching $\sim$2~kJ from the full 160-bunch train. Unfortunately, relatively little is known about how this energy is transported within the plasma but naïve assumptions can help frame the problem. For example, if all the energy deposited in the plasma is equally distributed to all plasma electrons and ions in the plasma source (assuming a diameter of 1~mm, which is typical at FLASHForward), the temperature of the plasma constituents will rise to $\sim$10~keV from a single bunch—sufficiently high to ionise almost all levels of argon. Furthermore, if no energy were to be lost from the plasma between bunches in the train via, for example, electromagnetic radiation or conduction by the plasma source, this temperature could rise to $\sim$1.6~MeV, orders of magnitude hotter than some fusion reactors. Lower temperatures may be reached by using cm-scale plasma-source diameters but only if the energy can be evenly distributed on a sufficiently short timescale. Ultimately, little is known about the operation of plasma accelerators at these temperatures, both from the point of view of the plasma physics and how they affect the wakefield process. For example, for a hot (100~eV) plasma with HALHF densities (\SI{6e14}{\per\cubic\cm}) the electron-ion collision time is on the order of microseconds--orders of magnitude longer than the temporal separation between bunches. It is therefore highly likely that the plasma will be far from equilibrium at the arrival time of the next bunch in the train, thus significantly modifying the wakefield properties. Targeted research, in the form of long-term PIC simulations, with all the necessary physics included, and experimental results, with direct plasma-electron and -ion temperature diagnostics, is planned at the University of Oxford.

\subsubsection{Cooling} 
The energy deposited in the plasma will make its way to the surrounding plasma vessel if no remedial measures are taken. Although the energy-transport channels in a plasma accelerator remain relatively unmapped, upper bounds on the required cooling rates can be calculated from the energy deposited in the plasma from the drive beam. The HALHF baseline parameters need around 16,000 bunches per second to achieve the necessary luminosity. With the 1.6~J/m of energy deposition previously calculated for a single acceleration event, the time-averaged cooling rate required for the entire bunch train corresponds to 38~kW/m, larger than the cooling rates expected at CLIC but on the same order of magnitude. Furthermore, if HALHF were to operate in a burst mode as originally proposed, the energy deposition over the bunch train would be 256~J/m in 2.56~{\textmu}s, likely leading to MPa principle stresses on the plasma source material. Novel plasma-source designs capable of withstanding the extreme and rapid stresses and temperature changes are therefore an important R\&D topic for HALHF.

\subsection{Damping rings}
\label{sec:dampingrings}

It is assumed that a polarized source with suitable performance and longevity will be available by the time of HALHF construction, allowing operation without an electron damping ring. If this turns out not to be the case, addition of a damping ring is a relatively minor perturbation on the design.

The positron damping-ring design for HALHF has unfortunately not been pursued due to lack of effort and no funding resources being available. The working assumption is that the design for the CLIC dual positron damping ring can be adapted to HALHF requirements. A detailed design for such a damping ring is a priority for the next stage in HALHF design, as it is well known that dampings rings can put severe limitations on bunch currents and the bunch-train pattern of the facility.

\subsection{Positron linac}
\label{sec:positronlinac}

The guiding principle of HALHF is to avoid the necessity for positron acceleration in plasmas \cite{Cao2024} by using metallic cavity structures to do this task. The asymmetric energy of HALHF is predicated on the requirement to miniminse the cost of this positron linac -- a symmetrical machine would require a positron linac to accelerate to half the CoM energy, which would be very expensive. Even with the current asymmetry, giving a positron energy of 41.7~GeV for the Higgs factory, the positron linac is a significant cost element and also is one of the longest parts of the complex. It is therefore important to minimise this cost as far as possible. An important element in such a minimisation is to maximise the accelerating gradient. This pushes us in the direction of a new technology, also being considered for a Higgs factory, that of cool copper~\cite{CCC_2023,CCC-EPSU2026}. 

As a new technology, we considered it important to choose a value of the gradient higher than achievable with warm linacs, such as that of the venerable SLAC linac, but well away from the state of the art. The Cool Copper collaborators have recently achieved 140~MV/m in beam tests at Next Linear Collider Test Accelerator (NLCTA) at SLAC, with a breakdown rate reduction by a factor of 50 compared to operation at room temperature~\cite{Nasr2021}. We propose a gradient of 40~MV/m, although this could well be safely increased in further design optimisations. 

\begin{figure}[ht]
    \centering
    \includegraphics[width=0.8\linewidth]{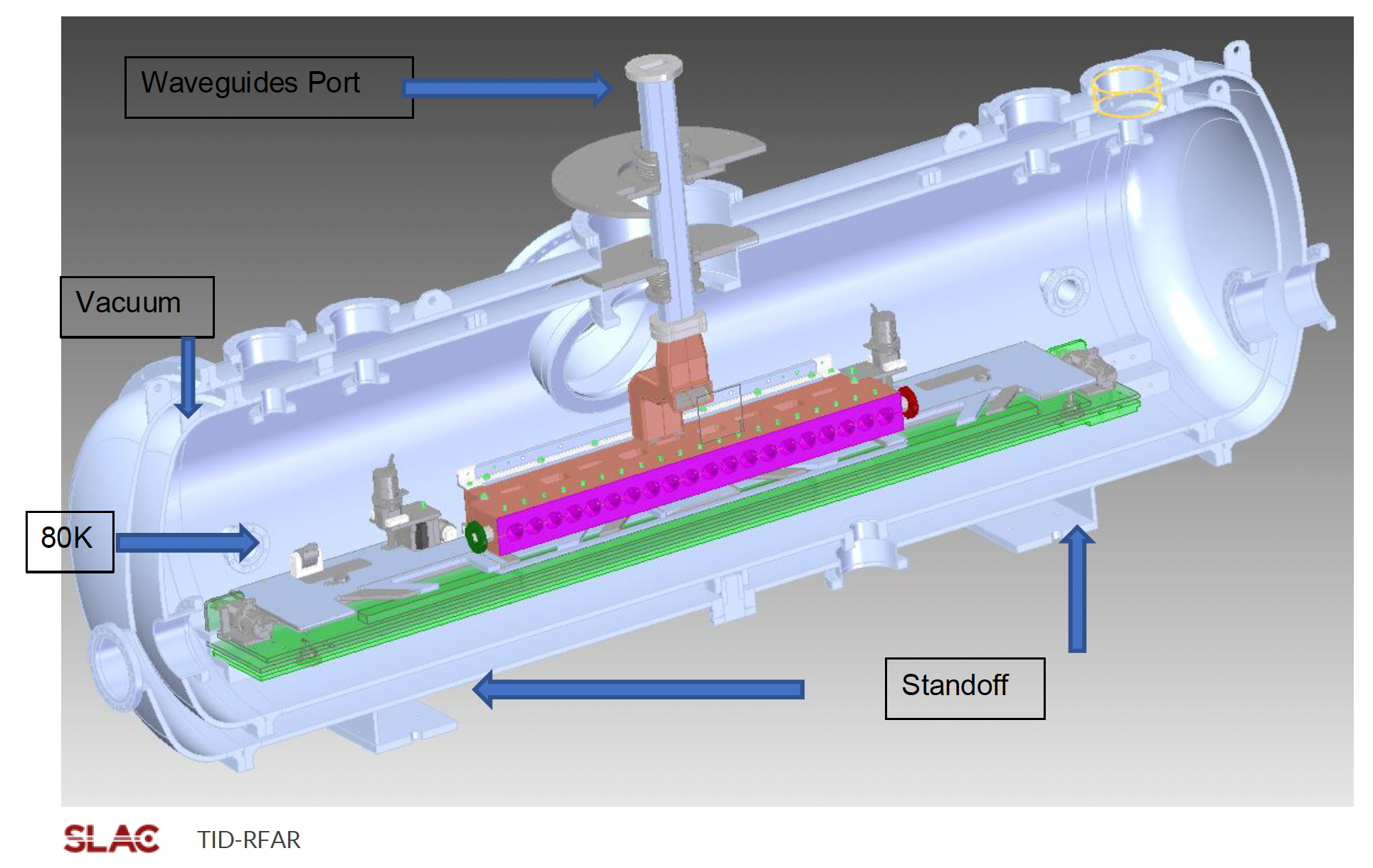}
    \caption{Schematic of the Quarter Cryo-Module to be tested at Argonne National Lab. The green ``raft'' structure supports the copper accelerating structures, shown in purple. From: C$^3$ Collaboration.}
    \label{fig:CC_QCM}       
\end{figure}

\begin{figure}[ht]
    \centering
    \includegraphics[width=\linewidth]{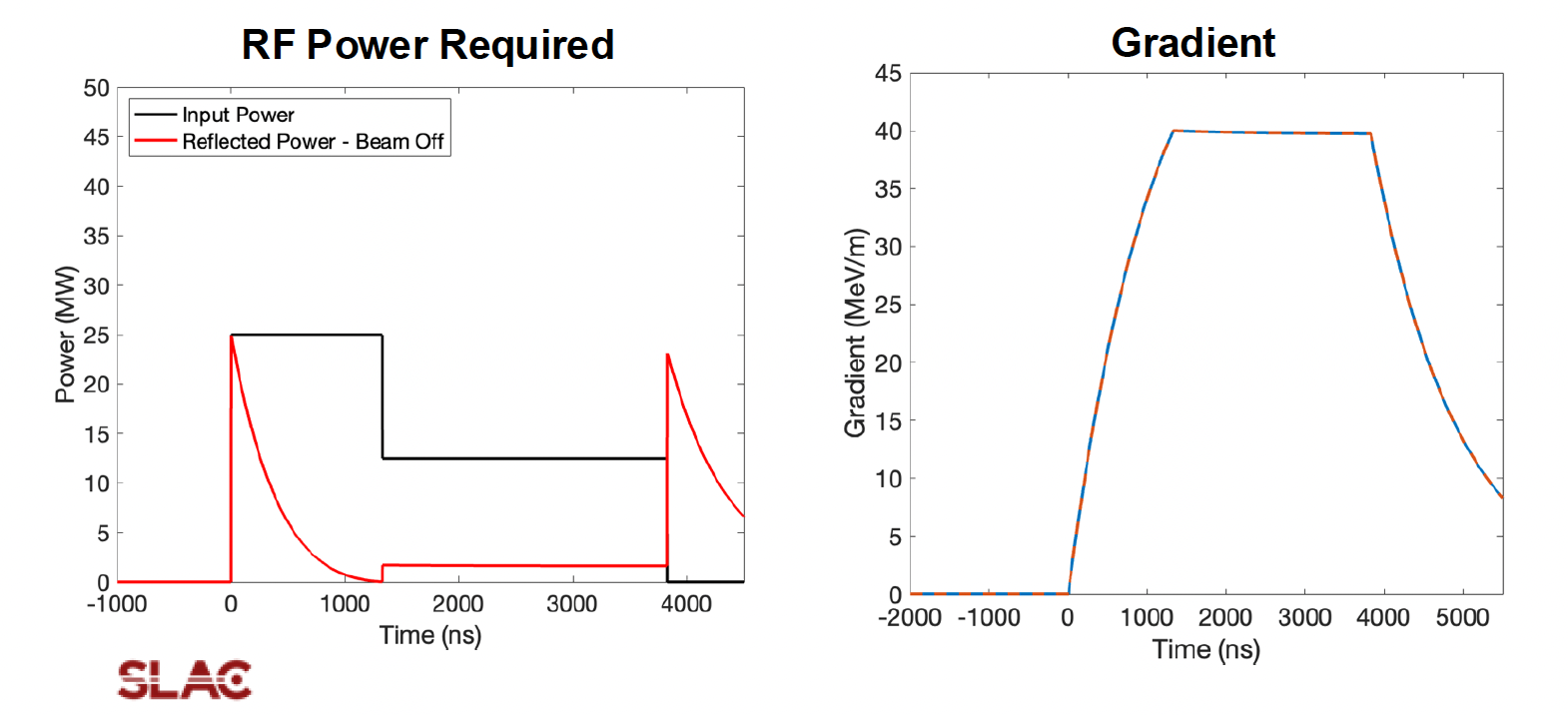}
    \caption{Power and gradient requirements for a cool copper linac with the HALHF bunch pattern.}
    \label{fig:CC_power}       
\end{figure}

We are collaborating closely with the Cool Copper researchers. They are currently
constructing a Quarter Cryo-Module with Rafts Assembly (see Fig.~\ref{fig:CC_QCM}) to be tested at high power at the Advanced Photon Source (APS) at Argonne
using the Linac High-Power RF Test stand and the Linac Extension Area. Tests will include:
\begin{itemize}
    \item a high power test measuring breakdown rate with a 30~MW klystron;
\item full power + beam tests;
\item a second phase with two structures and a 60 MW klystron;
\item a repeat full test with the quarter cryo-module.
\end{itemize}
Initial studies for a cool copper positron linac for HALHF indicate that the requirements should be readily achievable. Figure~\ref{fig:CC_power} shows the RF power required and the effective gradient for the HALHF bunch pattern.
If for whatever reason it transpires that cool copper does not provide a solution for HALHF's requirements, there is a fall-back, as shown in Fig.~\ref{Fig-2}, \textit{viz.} a warm linac with a gradient of 25~MV/m. Although this will obviously be longer and more expensive, it is rather similar to the SLAC linac, which typically operated with gradients between 17 and 20 MV/m. It is therefore a very conservative choice and a safe fall-back solution. More information is contained in the Cool Copper Collider submission to this process~\cite{CCC-EPSU2026}.

\subsection{Beam delivery system}
\label{sec:BDS}

The development of the beam delivery system (BDS) for HALHF is a crucial aspect of the overall collider design, ensuring efficient beam transport, focusing, and collimation. A more comprehensive overview of the BDS design and its fundamental principles can be found in the CLIC Project Implementation Plan (PIP) \cite{CLIC_PIP_2018}. This appendix provides a focused discussion on the specific challenges and limitations faced by the HALHF BDS, along with potential solutions and future plans.

\subsubsection{Adapting the CLIC BDS design to HALHF parameters}
The HALHF BDS team is currently addressing key constraints in the system’s design, primarily focusing on minimising the BDS length while preserving performance.
At this stage, the primary limitation of the HALHF BDS is the large emittance values, which lead to a significant reduction in luminosity at the IP. Unlike the more compact BDS systems in CLIC and ILC, HALHF's emittance constraints result in a longer and more complex BDS. To mitigate the impact of large emittance and optimise the BDS performance, several possible solutions are being explored:
\begin{itemize}
    \item optimising the optics: adjusting the chromaticity correction scheme and increasing dispersion through larger bending angles;
    \item expanding the collimation sections: ensuring efficient removal of beam halo to enhance beam quality;
    \item using advanced magnet technology: exploring Nb$_3$Sn superconducting magnets to reduce the system’s length constraints;
    \item reducing the beam emittance: shortening the required BDS length by operating closer to ILC/CLIC parameters, but at the cost of introducing an electron damping ring and making emittance preservation more challenging in the plasma linac.
\end{itemize}

\noindent
Recent studies underline that it will not be possible to fit the BDS within the desired length constraints unless we find a compromise between the plasma linac integration, which could alter beam dynamics and help in reducing emittance and adding a damping ring, which, while complex, might be the most effective way to shorten the BDS.

\subsubsection{Dual IP}

The dual BDS option, inspired by the CLIC dual IP design \cite{Cilento2021}, would further extend the overall system length. The additional separation beamline and dipoles required for the dual IP configuration increase complexity and engineering constraints. Given these challenges, the HALHF BDS length would need to exceed 2.6 km, with further extensions depending on the final implementation of the dual IP scheme.

\subsubsection{Future work}

To move towards a feasible BDS design, upcoming work will focus on refining the electron BDS and testing alternative optics solutions, assessing the feasibility of damping rings and their impact on reducing emittance and investigating plasma linac integration as a possible alternative to traditional acceleration schemes.
To evaluate dual IP implementation, considering both luminosity and length constraints will be necessary. Moreover, collaborating with plasma-accelerator experts to explore new approaches for collimation and beam focusing would be fundamental. 
\begin{table}[t]
\scriptsize
\vspace{-1cm}
    \begin{tabular}{p{0.6\linewidth}>
    {\centering}p{0.035\linewidth}>
    {\centering}p{0.05\linewidth}>
    {\centering}p{0.09\linewidth}>
    {\centering}p{0.085\linewidth}}%
    \hline
    {\bf R\&D element}&Start\\year&Duration\\(years) &Personnel\\(FTE years)& Capital\\(MCHF) \cr
    \hline
    \textit{Phase 1: Basic R\&D and integrated collider design} &  0 & 5 & &  \cr
    \hline
    Plasma accelerator R\&D:&  &  & &  \cr
    \tabindent Single-stage quality preservation at high efficiency demonstration (ongoing) & 0 & 5 & 90 & 13 \cr
    \tabindent Basic staging and beam-quality R\&D:&  &  & &  \cr
    \tabindent \tabindent Self-consistent PWFA staging simulations (incl. spin polarization) & 0 & 3 & 15 & 2 \cr
    \tabindent \tabindent Demonstrating achromatic staging optics (nonlinear plasma lens) & 0 & 3 & 5 & 3 \cr
    \tabindent \tabindent Quality-preserving stage-to-stage transport experiment & 3 & 2 &10 & 5\cr
    \tabindent Basic plasma heating and cooling R\&D&  &  & &  \cr
    \tabindent \tabindent Self-consistent long-term plasma evolution simulations & 0 & 3 & 10 & 1\cr
    \tabindent \tabindent Cooled plasma-cell development & 0 & 5 & 30 & 5 \cr
    \tabindent \tabindent High-peak-power plasma evolution experiment (in existing PWFA facilities) & 0 & 5 & 10 & 2 \cr
    Collider design (toward CDR): &  &  & &  \cr
    \tabindent Polarized $\rm{e}^+$ source R\&D & 0 & 5 & 16 & 3\cr
    \tabindent Positron linac design (e.g., cool copper) & 0 & 5 & 10 & 10 \cr
    \tabindent Drive-beam complex design (linac, combiner rings, etc.) & 0 & 5 & 20 & 7\cr
    \tabindent Beam-delivery system design (incl. double IP) & 0 & 5 & 10 & 1 \cr
    \tabindent Asymmetric physics and detector design (Not included in accelerator design) &  &  & & \cr
    \hline
    \textit{Phase 2: Key demonstrations} & 5 & 5 & &  \cr
    \hline
    Plasma demonstrations: & &  & &  \cr
    \tabindent Staging and stability demonstrator (new dedicated facility; SFQED application) & 5 & 5 & 25 & 60 \cr
    \tabindent High-average-power plasma-cell demonstration (upgrading an existing facility) & 5 & 5 & 20 & 20 \cr
    \tabindent Single-stage polarization preservation experiment & 7 & 3 & 15 & 20 \cr
    \tabindent Self-consistent full-train start-to-end simulations & 5 & 5 & 5 & 1 \cr
    Other systems demonstrations: & &  & &  \cr
    \tabindent Cool-copper RF linac demonstrator (parallel development to HALHF) &  &  & &  \cr
    \tabindent Polarized positron source demonstrator (parallel development to HALHF) &  &  & &  \cr
    \hline
    \textit{Phase 3: HALHF (all in one) demonstrator} & 10 & 5 & & \cr
    \hline
    Upgrade of staging facility with high-power plasma cells and RF & 10 & 5 & 20  & 25 \cr
    Upgrade to include increased beam quality & 12 & 3 & 20 & 25  \cr
    Upgrade to include spin polarized source & 13 & 2 & 10 & 10 \cr
    \hline
    \textbf{Total} &  & 15 yrs & 341 FTE yrs & 213 MCHF \cr
    \hline
    \end{tabular}
   \caption{HALHF R\&D Milestones and required resources to produce a Technical Design Report (TDR). (This table is reproduced from Table 2 in the Comprehensive Summary~\cite{Input2025}.)}
      \label{tab:2}
\end{table}

\begin{table}[!ht]
\tiny
    \begin{tabular}{>{\raggedright}p{0.17\linewidth}>
    {\centering}p{0.02\linewidth}>
    {\raggedright}p{0.08\linewidth}>
    {\centering}p{0.08\linewidth}>
    {\centering}p{0.10\linewidth}>
    {\centering}p{0.04\linewidth}>
    {\centering}p{0.05\linewidth}>
    {\raggedright}p{0.26\linewidth}}%
    \hline
Critical parameters                                               & TRL    & R\&D time (y) (design/total)                                                      & R\&D current (M€) & R\&D needed \\(M€; design/total)& FTE current & FTE-yrs needed & Comments                                 \cr
    \hline
Electron beams \textgreater{}~100~GeV & 1   & 7--8 / 11--13 &                                                                       & 7/100                                                & 1  & 40  & No PWFA-test facilities have produced \textgreater 100 GeV beams                                                      \cr
    \hline
Acceleration in one stage ($\sim$10~GeV)   & 5    & 5 / 9--10 &                                                                       & 10/100                                              & 3                                & 50                                                                   & AWAKE demonstration but  technology may not be suitable                                             \cr
    \hline
Plasma uniformity \\(long \& trans.)                                & 4      & 5/9--10    &                                                                       & 2/100                                                & 2                                                                       & 15                                                                   & AWAKE demonstration but  technology may not be suitable                                                              \cr
    \hline
Preserving beam quality/emittance                      & 3.5      & 7--8/11--13 & 0.5 (ERC + Oslo national)                                & 3 / 100                                              & 5                      & 25                                                                   & Normalized emittance preserved at \textless 3 um levels with small currents\cr
    \hline
Spin, polarization                                                & 2                                                                       & 5 / 9--10    &      0.1 (DESY)                                                                 & 3/100                                                & 1                                                 & 16                                                                   & Technology concept formulated                  \cr
    \hline
Stabilisation (active and passive)                                & 3                                                                        & 7--8 / 11--13 &                                                                       & 1/100                                               & 1                                                                       & 10                                                                   & Studies at AWAKE and LWFA, but not at HALHF requirements                                                          \cr
    \hline
Ultra-low-emittance beams                                         & 2                                                                           & 7--8 / 11--13&                                                                       & 3/100                                                & 0                                                                       & 20                                                                   & Not yet collider emittances; need better test facilities.                                                                                                                   \cr
    \hline
External injection and timing                                     & 4                                                                       & 7--8 / 11--13 &                                                                       & 1/100                                               & 0                                                                       & 10                                                                   & Precise timing for external injection demonstrated at AWAKE                                                                         \cr
    \hline
High rep-rate targetry with heat management                       & 2                                                                      & 5 / 9--10   &                                                                      & 7/100                                                & 3                                               & 40                                                                   & Heat modification of plasma properties/profile and  target cooling requires new concepts                 \cr
    \hline
Temporal plasma uniformity/stability                           & 4                                                                    & 5 / 9--10    &                                                                       & 3/100                                               & 0                                                                       & 10                                                                   & AWAKE demonstration but technology may not be suitable                        \cr
    \hline
Driver removal                                          & 2                                                                         & 7--8 / 11--13 &                                                                       & 2/100                                               & 0.5                                                                     & 10                                                                   & HALHF concept exists                                                     \cr
    \hline
Drivers @ high rep.~rate \& eff.& 5                                                                      & 5 / 9--10   &                                                                       & 5/100                                                & 0.5                                                                     & 10                                                                   & Similar to CLIC driver, demonstrated in CTF3         \cr
    \hline
Interstage coupling                              & 2                                                                & 7--8 / 11--13 & 1 (ERC)                                                        & 3/100                                                & 3                                                                       & 10                                                                   & HALHF concept exists              \cr
    \hline
Total system design with end-to-end                               & 3                                                                      & 3--4                                                       & ~0.5 (Oslo, Oxford)                                           & 3                                                     & 2                                                                 & 20                                                                   & Not yet at pre-CDR level. Aim for pre-CDR document early in 2026.    \cr
    \hline
Simulations                                                       & 5                                                                   & part of above                                                   & 0.5 (ERC + Oslo national)                                & part of above                                                                 & 4                                                 & 5                                                                    & Single-stage simulation well developed - dedicated framework (ABEL) for start-to-end                \cr
    \hline
Self-consistent design                                            & 4                                                                    & part of above                                                   & part of above                                                         & part of above                                                                 & in prev. 2 rows                                            & 5                                                                    & Plasma linac start-to-end simulations performed using HIPACE++/ABEL \cr          
     \hline
\end{tabular}
\caption{HALHF plasma-arm R\&D: Technology Readiness Levels (TRL), required resources and timescales to produce TDR. ``FTE current'' means currently in place; ``needed'' is integrated total requirement. ``R\&D current'' is only non-zero if dedicated to HALHF. Adapted from LDG Accelerator Roadmap review submission, February 2025. (This table is reproduced from Table 3 in the Comprehensive Summary~\cite{Input2025}.)}
      \label{tab:3}
\end{table}

\begin{figure}[!ht]
    \centering
    \includegraphics[width=0.62\textwidth]{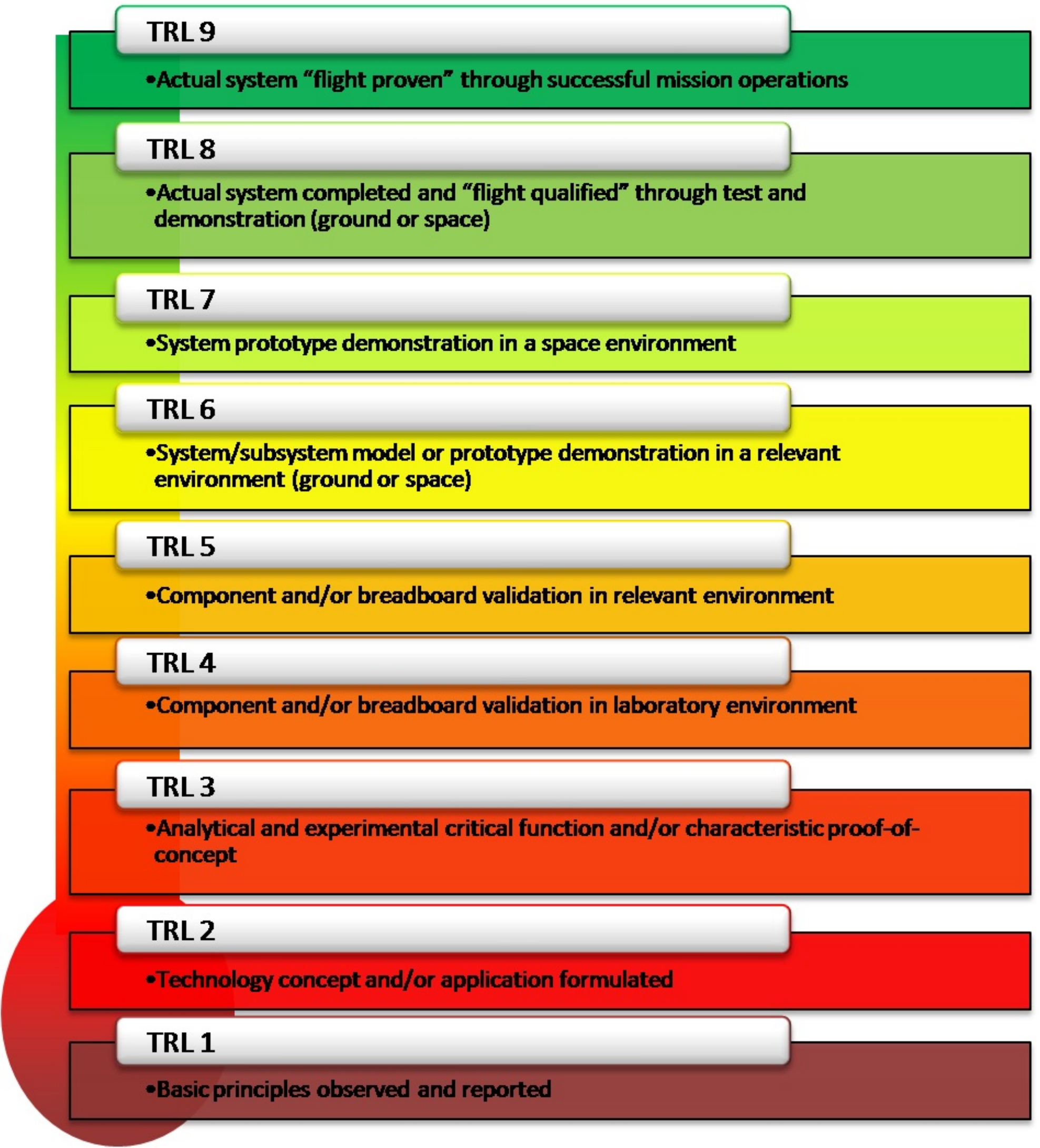}
    \caption{Definition of Technology Readiness Levels. Source:
NASA Procedures and Guidelines 7123.1 B.}
    \label{fig:trl}   
\end{figure}

\section{R\&D plan}
\label{sec:RandD}
In this section, we describe the R\&D necessary for the initial Phase-1 activities in some detail. Table~\ref{tab:2} summarises our estimate for the overall R\&D programme necessary to produce a Technical Design Report for HALHF, while Table~\ref{tab:3} breaks down the work for the PWFA arm in more granularity. The nine Technology Readiness Levels (TRL) for various subsystems and aspects of HALHF shown therein are defined in Fig~\ref{fig:trl}. 

While necessarily only approximate, we believe that funding close to what is indicated here (i.e., $\sim$5\% of the collider cost) would be sufficient to make a major step forward towards making a PWFA facility a reality for particle-physics applications. This in turn would have major implications for work towards a 10~TeV collider~\cite{10TeVESPP}, seen e.g.\ in the USA as the next step beyond an \ee\ Higgs factory.

A key and indeed minimum requirement for the approval of a technology of the novelty of HALHF is the production of a demonstrator that can exhibit the properties required of a user-driven facility. This demonstrator facility and the TDR are the main deliverables of the HALHF R\&D programme. The Phase 1 programme that starts in year 0 and lasts for 3--5 years, as shown in Table~\ref{tab:2}, can be described in somewhat more detail than the Phase~2 and 3 programmes that necessarily depend on success in Phase~1. 
\subsection{R\&D on single-stage properties}
The single-stage properties that constitute the first item in Phase 1 are a central theme of the world-wide PWFA research effort currently under way. The unique feature of HALHF research in this area is the requirement for high power in order to achieve the required luminosity. This is not being investigated at any significant level outside the HALHF Collaboration. Indeed, only the FLASHForward facility~\cite{Ashikin2016, Darcy2019} at DESY is capable of relevant experimentation using high-energy beams ($\geq 1$~GeV). The number of FTEs engaged in such experimentation, which is not explicitly for HALHF, is to the best of our knowledge of order three. A similar number is engaged on the design of plasma cells suitable for HALHF parameters; as is discussed in more detail below, a significant increase in numbers is required to make progress on the HALHF design. 
\subsection{R\&D on staging}
The R\&D on staging is currently planned using both LWFA and PWFA. The original staging experiment~\cite{Steinke2016} was carried out at the BELLA facility~\cite{BELLA} at Lawrence Berkeley National Laboratory and further work is planned there. While only low charge and low repetition rates are possible, this work is still relevant for HALHF. The PWFA staging activity is mostly under the aegis of the SPARTA~\cite{SPARTA} project, which although not specifically orientated towards HALHF, has a substantial overlap in personnel with the HALHF Collaboration and has goals that are very congruent. A dedicated HALHF activity would have significant synergetic effects and strengthen both projects towards the goal of producing a high-energy PWFA demonstrator. 

The optical elements required for staging are also being developed, again predominantly within the SPARTA project. First experimental tests of nonlinear plasma lenses, required for compact and achromatic beam transport, are currently being carried out at the CLEAR facility at CERN. This work needs to be expanded and a dedicated design team for HALHF is required to optimise the setup for the 48 HALHF stages in the current baseline design.

\subsection{R\&D on spin polarization}
There is currently little R\&D devoted to issues of spin-polarization preservation in PWFA. This aspect is a key advantage of linear colliders compared to circular colliders, so that it is essential to fund research in this area. Although PIC simulations imply that polarization is preserved during non-linear plasma acceleration, an experimental demonstration is necessary.

\subsection{R\&D on plasma heating and cooling}
R\&D activity related to plasma-cell design and cooling, a crucial topic for HALHF, aims to develop both software and hardware to study key thermodynamic processes at high-repetition-rate\cite{Darcy2022} and high-average-power plasma accelerators. The two main areas of work are firstly to understand how the energy deposited in the plasma evolves and secondly to assess how the temperature rise this produces modifies the plasma wakefields. 

Achieving these goals will require new approaches and solutions to elucidate the quantitative implications of high temperature on the plasma-wakefield process, enabling the development of strategies to combat plasma heating as well as techniques to cool high-average-power plasma stages. Usually, this is approached by simulation using PIC codes and then carrying out corresponding experiments. However, in the case of HALHF a different approach is needed because of the following challenges:
\begin{itemize}
\item PIC codes only simulate the physics over timescales orders-of-magnitude shorter (fs--ps) than the separation between high-repetition-rate bunches at HALHF (ns--{\textmu}s). They therefore omit aspects of physics required to simulate a high-repetition-rate plasma accelerator. Even were these processes to be contained in the PIC codes, simulations over such a long timescale are prohibitively expensive computationally because of the nm-level resolution at which important plasma-physics processes take place;
\item Experiments typically use measurements of the plasma-accelerated beam to infer the instantaneous plasma properties. Unfortunately, plasma temperature changes manifest themselves in such measurements in similar ways to e.g.\ changes in plasma density. This makes such indirect diagnostics unsuitable to measure the thermodynamic changes in the plasma.
\end{itemize}

The necessary R\&D must be tackled both in the laboratory and in a beam line. A HALHF laboratory to pursue design of suitable plasma cells must be established at a suitable site. A beam-line to measure the properties of such cells would be located at a PWFA laboratory. The most suitable currently operating is FLASHForward at DESY~\cite{Darcy2019} but FACET-II at SLAC~\cite{FACETII}, or ~EuPRAXIA~\cite{Assmann2020} (depending on the timescale for funding), could also be possible sites. Detailed assessments of necessary resources for this programme are currently being completed, informing similar estimates for the other aspects of the HALHF R\&D programme.
\subsection{R\&D on overall collider design and non-plasma subsystems}
The next items in Table~\ref{tab:2} refer to the overall collider R\&D, which would aim to produce a full Conceptual Design Report for HALHF within five years of sufficient funding being allocated. 

There is significant activity in the design of the positron source, which is also required for ILC, as discussed in Sect.~\ref{sec:positronsource}. However, this is insufficient to provide a CDR for HALHF, which would require not only several FTEs of sophisticated engineering design but also capital expenditure to produce prototypes and testing of the necessary rotating target. 

For the cool copper linac, we hope that a small HALHF effort here will be sufficient to ``piggy-back'' on the effort centered at SLAC to develop this technology. We are in close contact with the C\textsuperscript{3} Collaboration~\cite{CCC-EPSU2026} on the design of a linac suitable for the rather conservative HALHF parameters. The necessary design work and help in the development of the technology nevertheless requires dedicated HALHF staff to interface with the C\textsuperscript{3} effort.

The drive-beam complex is very similar to that of CLIC but nevertheless has distinctive features that will require dedicated design work in close collaboration with our CLIC colleagues. This work builds on the many FTEs of effort so far dedicated to CLIC.   

The ongoing design for the BDS is described in Sect.~\ref{sec:BDS}. The current design effort leverages the work done on the CLIC BDS and is dedicated to HALHF. Current funding will however expire within two years and this must be replaced and enhanced if the design is to be successful. The large emittance of the electron beam leaving the PWFA accelerator causes significant difficulties compared to the much smaller assumed emittance in the CLIC design.
\subsection{R\&D on particle-physics experimentation}
The design of detectors suitable for HALHF is a significant task; the current effort has so far only been able to scratch the surface. Not only will some dedicated R\&D be required for specialised detectors in the forward direction and for luminosity monitoring, but also significant software effort is required to modify \textit{GEANT} to deal with asymmetric collisions. However, the required effort is not part of the current programme described in this document, which is dedicated to accelerator-related topics.
\subsection{Currently missing R\&D activities}
In addition to the topics explicitly pulled out in Tables~\ref{tab:2} and \ref{tab:3}, there are significant other topics where no HALHF design work is currently being undertaken due to insufficient resources being available. The most important of these is the damping rings. However, we are hopeful that some HALHF collaborators will be able to begin this research in the next few months. It is expected that overlap at the interfaces with the linac and positron-source groups will provide sufficient additional effort to allow a damping-ring scheme to be devised at least at a conceptual level. Work on producing beam dynamics for some of the HALHF components has begun but must be greatly strengthened as e.g.\ details of the positron source cannot be finalised without a full HALHF beam lattice.

\newpage
\noindent

\bibliographystyle{elsarticle-harv} 

\end{document}